\documentclass[pra,preprint,superscriptaddress,showpacs]{revtex4}

\usepackage{times}
\usepackage{amsbsy,amssymb}
\usepackage{graphicx,color}

\def\one{{\mathchoice {\rm 1\mskip-4mu l} {\rm 1\mskip-4mu l} {\rm
1\mskip-4.5mu l} {\rm 1\mskip-5mu l}}}
\def\beqa{\begin{eqnarray}}
\def\eeqa{\end{eqnarray}}

\newcommand{\ket}[1]{|{#1}\rangle}
\newcommand{\bra}[1]{\langle{#1}|}

\newcommand{\cmplxs}{{\mathbb C}}

\newcommand{\rlxs}{{\mathbb R}}
\newcommand{\zlxs}{{\mathbb Z}}

\newcommand{\mComment}[1]{}
\newcommand{\gComment}[1]{}
\newcommand{\lComment}[1]{}
\newcommand{\hComment}[1]{}
\newcommand{\rComment}[1]{}
\renewcommand{\mComment}[1]{\textcolor{blue}{Manny: #1}}
\renewcommand{\gComment}[1]{\textcolor{red}{Gerardo: #1}}
\renewcommand{\lComment}[1]{\textcolor{green}{Lorenza: #1}}
\renewcommand{\hComment}[1]{\textcolor{magenta}{Howard: #1}}
\renewcommand{\rComment}[1]{\textcolor{purple}{Rolando: #1}}

\def\fu{\mathfrak{u}}

\def\fc{\mathfrak{c}}

\def\fh{\mathfrak{h}}

\def\fr{\mathfrak{r}}

\def\fsu{\mathfrak{su}}

\def\fso{\mathfrak{so}}

\begin{document}
\title{Nature and Measure of Entanglement in Quantum Phase Transitions}

\author{Rolando Somma}
\thanks{Corresponding author. Email: somma@lanl.gov}
\affiliation{Los Alamos National Laboratory, Los Alamos, NM 87545, USA }
\affiliation{Centro At\'omico Bariloche and Instituto Balseiro,
8400 San Carlos de Bariloche, Argentina}

\author{Gerardo Ortiz}
\affiliation{Los Alamos National Laboratory, Los Alamos, NM 87545, USA }

\author{Howard Barnum}
\affiliation{Los Alamos National Laboratory, Los Alamos, NM 87545, USA }

\author{Emanuel Knill}
\thanks{Current address: National Institute of Standards and Technology,
Boulder, CO 80305, USA}
\affiliation{Los Alamos National Laboratory, Los Alamos, NM 87545, USA }

\author{Lorenza Viola}
\thanks{Current address: Department of Physics and Astronomy, Dartmouth 
College, Hanover, NH 03755, USA}
\affiliation{Los Alamos National Laboratory, Los Alamos, NM 87545, USA }

\date{\today}

\begin{abstract}
Characterizing and quantifying quantum correlations in states of
many-particle systems is at the core of a full understanding of phase
transitions in matter.  In this work, we continue our investigation of
the notion of generalized entanglement [Barnum {\it et al.}, Phys.
Rev. A {\bf 68}, 032308 (2003)] by focusing on a simple Lie-algebraic
measure of purity of a quantum state relative to an observable set.
For the algebra of local observables on multi-qubit systems, the
resulting local purity measure is equivalent to a recently introduced
global entanglement measure [Meyer and Wallach, J. Math. Phys. {\bf
43}, 4273 (2002)].  In the condensed-matter setting, the notion of
Lie-algebraic purity is exploited to identify and characterize the
quantum phase transitions present in two exactly solvable models: the
Lipkin-Meshkov-Glick model, and the spin-1/2 anisotropic XY model in a
transverse magnetic field.  For the latter, we argue that a natural
fermionic observable-set arising after the Jordan-Wigner
transformation, better characterizes the transition than alternative
measures based on qubits. This illustrates the usefulness of going
beyond the standard subsystem-based framework while providing a global
disorder parameter for this model.  Our results show how generalized
entanglement leads to useful tools for distinguishing between the
ordered and disordered phases in the case of broken symmetry quantum
phase transitions.  Additional implications and possible extensions of
concepts to other systems of interest in condensed matter physics are
also discussed.
\end{abstract}

 \columnseprule 0pt

\maketitle

\section{Introduction}
\label{intro}
Quantum Phase Transitions (QPTs) are qualitative changes occuring in
the properties of the ground state of a many-body system due to
modifications either in the interactions among its constituents or in
their interactions with an external probe \cite{sa}, while the system
remains at zero temperature.  Typically, such changes are induced as a
parameter $g$ in the system Hamiltonian $H(g)$ is varied across a
point at which the transition is made from one quantum phase to a
different one.  Often some correlation length diverges at this point,
in which case the latter is called a {\em quantum critical point}.
Because thermal fluctuations are inhibited, QPTs are purely driven by
quantum fluctuations: fluctuations or correlations in the value of
some observable or observables that occur in a pure state.  Thus, these 
are purely quantum phenomena: a classical system in a pure state cannot
exhibit correlations. Prominent examples of QPTs are the quantum
paramagnet to ferromagnet transition occurring in Ising spin systems
under an external transverse magnetic field \cite{liscma,pf,bamc}, the
superconductor to insulator transition in high-$T_c$ superconducting
systems, and the superfluid to Mott insulator transition originally
predicted for liquid helium and recently observed in ultracold atomic
gases \cite{greiner}.

Since entanglement is a property inherent to quantum states and
intimately related to quantum correlations \cite{sc}, one would expect
that, in some appropriately defined sense, the entanglement present in
the ground state undergoes a substantial change across a point where a
QPT occurs. Recently, several authors attempted to better understand
QPTs by studying the behavior of different measures of entanglement in
the ground state of exactly solvable models (see
\cite{osni,osamfafa,arbove,vidal,baknorsovi,vers} for representative
contributions).  Such investigations primarily focused on
characterizing entanglement using information-theoretic concepts, such
as the entropy of entanglement \cite{bedismwo} or the concurrence
\cite{wo}, developed for {\em bipartite} systems. In particular, a
detailed analysis of the two-spin concurrence has been carried out for
the XY model in a transverse magnetic field \cite{osni,osamfafa},
whereas the entanglement between a block of nearby spins and the rest
of the chain has been considered in \cite{vidal}.  While a variety of
suggestive results emerge from such studies, in general a full
characterization of the quantum correlations near and at a quantum
critical point will not be possible solely in terms of bipartite
entanglement.  Identifying the entanglement measure or measures that
best capture the relevant properties close to criticality, including
the critical exponents and universality class of the transition,
remains open problems in quantum information and condensed matter
theory.

In Refs. \cite{baknorvi,baknorsovi}, we introduced {\it Generalized
Entanglement} (GE) as a notion extending the essential properties of
entanglement beyond the conventional subsystem-based framework.  This
notion is general in the sense that it is definable relative to {\em
any distinguished subset of observables}, without explicit reference
to subsystems, which makes it directly applicable to any algebraic
language used to describe the system (fermions, bosons, spins, etc.)
\cite{baor,advances,eckert}.  Founding the notion on a distinguished
set of observables makes it especially well suited to studying QPTs,
as our definition makes the existence of GE equivalent to the
existence of nonzero correlations or fluctuations in those
observables. The basic idea is that any quantum state gives rise to a
reduced state on the distinguished subset of observables \cite{Note0}.
These reduced states form a convex set; as with standard quantum
states, there are pure (extremal) and mixed (non-extremal)
ones~\cite{Note1}. We define a {\em generalized entangled} pure state,
relative to a subspace of observables, to be one whose reduced state
on that subspace is {\em mixed}. Although we will have little occasion
in the present context to apply it to states that are mixed relative
to the full set of observables, we extend this notion to include mixed
states by defining a generalized entangled mixed state to be one that 
cannot be written as a convex combination of generalized-unentangled 
pure states.

The special case in which the observable set is a Lie algebra is often
important in Physics.  In a broad class of such algebras described
below, the algebra is not only a subspace of operators, but is such
that we can define a natural Hermitian projection onto that subspace.
Then a simple (global) measure of GE for quantum states is provided 
by what we call the {\em purity relative to the algebra}.  This is
defined as the squared length of the projection of the Hermitian
operator (density matrix) representing the state onto the algebra.  
As argued in \cite{baknorsovi}, if the correct algebra is chosen, the
purity contains information about the relevant quantum correlations
that uniquely identify and characterize QPTs of the system.

In this paper, we deepen and expand the analysis initiated in
\cite{baknorsovi}, by focusing on the detection of QPTs due to a
broken symmetry as revealed by the behavior of an appropriate relative
purity of the ground state.  In Section \ref{sectionpurity}, we recall
the relevant mathematical setting and the definition of the relative
purity as a function of the expectation values of the distinguished
observables.  In Section \ref{examples}, we discuss several examples
where the relative purity is seen to provide a natural measure of
entanglement. In Section \ref{purity-QPT}, we illustrate some physical
criteria that are relevant in choosing the appropriate observable
subalgebra and using GE as an indicator of QPTs.  In Sections
\ref{lmg} and \ref{xymodel} we explicitly characterize the QPTs
present in the so-called Lipkin-Meshkov-Glick (LMG) model
\cite{limegl,orsoduro} and in the one-dimensional spin-1/2 anisotropic XY model
in a transverse magnetic field, respectively.  This is done by
studying the properties of the purity relative to different algebras
of observables in the ground state of both models.  We find the
relevant critical exponents for these models, and in the case of the
anisotropic XY model in a transverse magnetic field, obtain a new
``global'' disorder parameter, the variance of the number of spinless
fermionic excitations in a Jordan-Wigner-transformed representation of
the system.  Finally, we provide in separate Appendices the details
underlying various statements made in the main body of the paper.
These include the relationship between standard separability and GE
(Appendix A), the GE properties of two special classes of spin states,
the cluster and valence bond solid states (Appendix B), the proof of
the relationship between the local purity and the Meyer-Wallach
entanglement measure (Appendix C), and the semiclassical properties of
the LMG model in the thermodynamic limit (Appendix D).

\section{Generalized entanglement and relative purity}
\label{sectionpurity}

In the GE approach, entanglement is considered as an {\em
observer-dependent} property of a quantum state, which is determined
by the physically relevant point of view through the expectation
values of a distinguished subset of observables. Whenever a preferred
decomposition into subsystems is specified in terms of an appropriate
(physical or encoded \cite{klv,defilippo,vkl,zanardi}) tensor product
structure, GE becomes identical to standard entanglement provided that
distinguished observables corresponding to all {\em local} actions on
the individual subsystems are chosen: in particular, for ${\cal H}=
\otimes_i {\cal H}_i$ with dim(${\cal H}_i)=d_i$, standard
entanglement of states in ${\cal H}$ is recovered as GE relative to
$\fh_{loc}= \oplus_i \fsu(d_i)$ \cite{baknorvi,baknorsovi} (see also
Appendix \ref{app1}). In fact, the subsystems relative to which
standard entanglement is defined (whether directly identifiable with
physical degrees of freedom or related to ``encoded'' or ``virtual''
ones) are always understandable in terms of appropriate (associative)
algebras of local observables.  This has been observed before, e.g. in
\cite{vkl,zanardi} (see also \cite{zll} for a recent analysis).
However, it is important to realize that the GE notion genuinely
extends the standard entanglement definition, and does not coincide
with or reduce to it in general.  On one hand, this may be appreciated
by noticing that even for situations where a subsystem partition is
naturally present, states which are manifestly {\em separable}
relative to such a partition may possess GE relative to an algebra
different from $\fh_{loc}$ (see the two spin-1 example of Section
\ref{examples}). On the other hand, as also emphasized in
\cite{baknorsovi}, GE is operationally meaningful in situations where
{\em no} physically accessible decomposition into subsystems exists,
thus making conventional entanglement not directly definable.

\subsection{Relative purity for faithfully represented Lie algebras} 

As mentioned in the Introduction, we will focus on the case where the
distinguished observables form a Lie algebra $\fh$ of linear
operators, acting on a finite-dimensional state space ${\cal H}$ for
the system of interest, ${\cal S}$. (Note that we will not usually
distinguish between the abstract Lie algebra isomorphic to $\fh$, and
the concrete Lie algebra $\fh$ of operators that faithfully represents
it on ${\cal H}$.)  We will assume $\fh$ to be a real Lie algebra
consisting of Hermitian operators, with the bracket of two linear
operators $X$ and $Y$ being given by \beqa [X,Y] = i (XY - YX)\;.
\eeqa In this way, operators in $\fh$ can be directly associated with
physical observables. For the same reason, we will also use a slightly
nonstandard (but familiar to physicists) notion of the Lie group
generated by $\fh$, involving the map $X \mapsto e^{iX}$ instead of
the mathematicians' $X \mapsto e^X$, for $X \in \fh$.  No assumption
that the Lie algebra acts irreducibly on ${\cal H}$ (i.e., that it
admits no nontrivial invariant subspaces) will be made, but important
consequences of making such an assumption will be discussed.  We will
also assume the Lie algebra to be closed under Hermitian conjugation.
This implies that it is a {\em reductive} algebra (not to be confused
with reducibility of the representation).  In our context, a reductive
Lie algebra is best thought of as the product (direct sum, as a vector
space) of a finite number of {\em simple} Lie algebras, and a finite
number of copies of a one-dimensional Abelian Lie algebra.  A {\em
simple} Lie algebra is a non-Abelian one possessing no nontrivial
ideals, where an ideal is a subalgebra invariant under commutation
with anything in the algebra; the relevant property of ideals here is
that they can be quotiented out of the algebra, allowing it to be
written as a nontrivial product of ideal and quotient; thus simple Lie
algebras are non-Abelian ones that cannot be decomposed into factors,
so the factorization used in defining reductive Lie algebras above is
maximal.  The product (direct vector-space sum) of a finite number of
simple Lie algebras is called {\em semisimple}, and thus a reductive
algebra is the product of a semisimple and an Abelian part.  The
reader is referred to \cite{georgi,hall,humphreys,fh} for relevant
background on Lie algebras and their representation theory.  As this
subsection unfolds, we will summarize much of this representation
theory in a way suited to our needs, and the reader should concentrate
on understanding the content of the statements, and not vex him or
herself unduly about understanding why they are true.

We will consider pure quantum states of ${\cal S}$, $\ket{\psi} \in {\cal H}$, 
as well as mixed quantum states of ${\cal S}$,  described by density matrices 
$\rho$ acting on ${\cal H}$.  Since
$\fh$ was assumed closed under Hermitian conjugation,
the projection of a quantum state 
$\rho$ onto $\fh$ with respect to the trace inner product is 
uniquely defined.  Let  ${\cal P}_{\fh}$ denote the projection map, $\rho
\mapsto {\cal P}_{\fh}(\rho)$.  
If $\rho$ is a pure state, $\rho=\ket{\psi}\bra{\psi}$, the {\em purity
of  $\ket{\psi}$ relative to $\fh$} (or $\fh$-purity) is defined as the
squared length of the  projection according to the trace inner product norm
\cite{baknorvi}; that is 
\beqa P_{\fh} ( \ket{\psi} ) 
={\sf Tr}[({\cal P}_{\fh}(\ket{\psi}\bra{\psi} ))^2]\;.
\label{p0}
\eeqa 
The $\fh$-purity may be explicitly evaluated upon selecting an operator
basis  ${\cal B}=\{ A_1, \ldots, A_L \}$ for $\fh$. By assuming the
$A_\alpha$ to be Hermitian, 
\begin{equation}
A_\alpha=A_\alpha^{\dagger} \;,
\end{equation}
and orthogonal, 
\begin{equation}
\label{normalise}
{\sf Tr} (A_\alpha A_\beta)=\delta_{\alpha,\beta} \;,
\end{equation}
Eq. (\ref{p0}) may be rewritten as 
\begin{equation}
\label{purity0}
P_{\fh} ( \ket{\psi} ) = {\sf Tr}\Big[
\sum\limits_{\alpha,\beta=1}^L {\sf Tr}(A_\alpha \rho) {\sf
Tr}(A_\beta \rho) A_\alpha A_\beta \Big] = \sum \limits_{\alpha=1}^L \langle
A_\alpha \rangle^2 \;,
\end{equation}
where $\langle A_\alpha \rangle$ denotes the expectation value of the
observable $A_\alpha$ in the pure state $\ket{\psi}$. 

An important property following is that the
$\fh$-purity  is {\em invariant} under group transformations: if a new
basis for $\fh$ is introduced by  letting $\tilde{A}_\alpha= D^\dagger
A_\alpha D$, with $D=\exp({i \sum\limits_{\beta=1}^L  t_\beta
A_\beta})$, $D^\dagger D=\openone$,  and $t_\beta$ real numbers, then
one finds
\begin{equation}
\tilde{P}_{\fh} ( \ket{\psi} ) = 
\sum\limits_{\alpha=1}^L \langle \tilde{A}_\alpha \rangle ^2 =
\sum\limits_{\alpha=1}^L \langle A_\alpha \rangle ^2 =
{P}_{\fh} ( \ket{\psi} )\; .
\label{pinvariance}
\end{equation}
Sometimes it is useful to introduce a common normalization factor ${\sf
K}$  in order to set the maximum value of the purity to 1, in which
case Eq. (\ref{purity0})  becomes
\begin{equation}
\label{purity1}
P_{\fh}( \ket{\psi}) = {\sf K} \sum 
\limits_{\alpha=1}^L \langle A_\alpha \rangle^2.
\end{equation}

As mentioned earlier, a pure quantum state $\ket{\psi}$ is defined to
be {\it generalized entangled} ({\it generalized unentangled})
relative to $\fh$ if it induces a mixed (pure) state on that set of
observables. When $\fh$ is a complex semisimple Lie algebra acting
{\em irreducibly} on ${\cal H}$, it was shown in \cite{baknorvi}
(Theorem 14, part (4)) that $\ket{\psi}$ is generalized unentangled
with respect to $\fh$ {\it if and only} if it has maximum $\fh$-purity, 
and generalized entangled otherwise.  Under the same assumptions, the
abovementioned Theorem (part (3)) also leads to the identification of
the generalized unentangled pure states as the {\it generalized
coherent states} (GCSs) associated with $\fh$ \cite{gi,pe,zhfegi}.  In
other words, all generalized unentangled states are in the (unique)
orbit of a minimum weight state of $\fh$ (taken as a reference state)
under the action of the Lie group.  Remarkably, GCSs are known to
possess {\em minimum invariant uncertainty}, $(\Delta F)^2
(\ket{\psi})= \sum_\alpha \Big[ \langle A_\alpha^2 \rangle - \langle
A_\alpha \rangle ^2 \Big]$ \cite{del1,del2}, so that, similar to the
familiar harmonic-oscillator ones, they may be regarded in some sense
as closest to ``classical" states.

Our characterization theorem for generalized unentangled states on
irreducible representations used some standard facts from the theory
of semisimple Lie algebras and their representations that will also be
useful in the discussion of reducible representations in the next
subsection.  These are the existence of Cartan (in the semisimple
context, maximal Abelian) subalgebras, their conjugacy under the
action of the Lie group associated with the algebra, and the fact that
any finite-dimensional representation, given a choice of Cartan
subalgebra (CSA), decomposes into mutually orthogonal ``weight
spaces,'' which are simultaneously eigenspaces of all CSA elements.
The map from CSA elements to their eigenvalues on a given weight space
is a linear functional on the CSA called the ``weight'' of that weight
space.  The theorem also uses the observation that the projection of
the state into the Lie algebra is necessarily a Hermitian element of
that algebra, hence semisimple (diagonalizable), hence belonging to
some CSA, which we call its {\em supporting CSA}.  Frequently, semisimple
Lie algebras are presented by giving a {\em Cartan-Weyl} basis, consisting
of a set of commuting, jointly diagonalizable 
operators that generate a CSA of the algbera,
and a set of so-called ``Weyl operators'' that are non-diagonalizable, 
and act to take a state in one weight space to a state in another (or
else annihilate it):  in physical examples these are often called
``raising and lowering operators.''
Normalized states
correspond to normalized linear functionals on the Lie algebra; when a
Cartan-Weyl basis for the algebra is chosen such that the CSA
distinguished by the basis is the supporting CSA for a given state,
the state is zero except on the CSA part of the basis.  On the CSA,
the state is some convex combination of the weights, that is an
element of the {\em weight polytope} (which is defined as the convex
hull of the weights).  So it turns out that extremal states on the Lie
algebra correspond to extremal points of the weight polytope.  This 
applies regardless of whether the representation is irreducible
or not.  For irreducible representations (irreps), the extremal points
of the weight polytope are also highest-weight states of the irrep.
Reducible representations are discussed in the next subsection (along
with some comments on reductive algebras).

In preparation for that, we introduce another aspect of standard Lie
theory: the Weyl group. Besides being able to take any CSA to any
other CSA, the Lie group also acts on the weight polytope for a given
CSA, by reflections in a set of hyperplanes through the origin.  The
group these generate is called the Weyl group.  Considered together,
the hyperplanes divide the weight space into a set of convex cones,
sometimes called {\it Weyl chambers}, whose points are at the origin,
and whose union with the hyperplanes is the entire space.  Any such
cone can be mapped to any other via the Weyl group action, and the
weight polytope of the representation is the convex hull of the Weyl
group orbits of the weights in the closure of any single Weyl chamber.

\subsection{Irreducibly vs reducibly represented Lie algebras} 

It is important to realize that the relationships just mentioned
between maximal purity, generalized coherence, and generalized
unentanglement established for a pure state relative to an irreducibly
represented algebra $\fh$ do {\em not} automatically extend to the
case where $\fh$ acts reducibly on ${\cal H}$.  We will discuss
semisimple algebras first and then, because the algebra we use to
analyze the LMG model is Abelian, the case of reductive algebras.

If $\fh$ is semisimple, a generic finite-dimensional representation of
$\fh$ may be decomposed as a direct sum of irreducible invariant
subspaces, ${\cal H} \simeq \oplus_\ell {\cal H}_\ell$, with each of
the ${\cal H}_\ell$ being in turn the direct sum of its weight spaces.
Every irrep appearing in the decomposition has a highest (or lowest)
weight, and for each of these irreps, there is a manifold of GCSs for
the irrep constructed as the orbit of a highest weight state for that
irrep.  The weight polytope for the {\em reducible} representation
will be the convex hull of those for all the irreps contained in it.
Because of this, the GCSs for these irreps will {\em not}, in general,
all satisfy the extremality property that defines generalized
unentangled states.  This reflects the fact that even for a state
belonging to a specific $\fh$-irrep, GE is a property which depends in
general on how the state relates to the whole representation, not
solely the irrep.  Nor is there necessarily a single weight, for one
of the constituent irreps, that generates (as the convex hull of the
Weyl group orbit) the weight polytope of the reducible representation.
Indeed, the extremal weights in the weight polytope, which correspond
to generalized unentangled states, need not all have the same length.
Since this squared length is the $\fh$-purity (as defined in
Eq. (\ref{purity0})) of the corresponding state, it is thus no longer
the case that all generalized unentangled states have maximal
Lie-algebraic purity.  However, maximal purity remains a {\em
sufficient}, though no longer a necessary, condition for generalized
unentanglement.  If the algebra is reductive, the expectations of a
maximal commutative subalgebra now include ones for the Abelian part
of the algebra, i.e. operators that commute with the entire algebra.
These must be proportional to the identity on each irrep, but may have
different eigenvalues (possibly degenerate) on different irreps.
States on this algebra then involve not just weights for the
semisimple part of the algebra, but expectation values for the Abelian
part of the algebra as well.  These can distinguish different subsets
of the irreps, and so irreps whose highest weight for the semisimple
part is not extremal for the semisimple part, may become extremal
(generalized unentangled) in the full reductive algebra.  However,
maximal quadratic purity will remain a sufficient, though in general
still not necessary, condition for a state being generalized unentangled.

More intuition about GE, purity, and GCSs may be gained from simple
examples.  Consider a physical system which is composed of two
spin-$1/2s$ (namely, two qubits), and let them be labeled by $A, B$,
with ${\cal H} = {\cal H}_A \otimes {\cal H}_B = \cmplxs^4$, and
corresponding $\fsu(2)$ generators $\sigma_\alpha^A$,
$\sigma_\alpha^B$, $\alpha \in \{x,y,z \}$.  Consider GE relative to a
{\em global} representation of $\fsu(2)$, whose total-spin generators
are $J_\alpha= \sigma_\alpha^A + \sigma_\alpha^B$.  This
representation splits into two irreps, the one-dimensional singlet
representation with $J=0$ and the three-dimensional triplet
representation with $J=1$.  The generalized unentangled states
relative to this representation of $\fsu(2)$ are those for which there
exists an $\alpha$ such that the state is a $\pm 1$ eigenstate of
$J_\alpha$.  With respect to the CSA $\fc= \{J_z \}$, those are the
states $\ket{\uparrow, \uparrow}, \ket{\downarrow,\downarrow}$, which
are also GCSs (with purity equal to 1).  No generalized unentangled
state is contained in the singlet irrep.  In particular, neither the
spin-zero state in the triplet, nor that which spans the singlet, are
generalized unentangled (they both have purity equal to 0), nor are
they on highest-weight orbits (thus GCSs). 

As another example consider a single spin-1 system, whose state space
${\cal H}=\cmplxs^3$ carries an irrep of $\fsu(2)$ \cite{baknorsovi}.
In this case, for any choice of spin direction (say $z$) only the $J_z
= \pm 1$ eigenstates are generalized coherent.  There is also a
one-dimensional $J_z = 0$ eigenspace.  The maximal-purity states are
also the highest-weight states; however, the pure $J_z=0$ eigenstate
is not a GCS, has zero purity, and is generalized entangled.  If, for
the same system, a distinguished algebra $\fso(2)$ generated by $J_z$
alone is chosen, then the representation reduces as the direct sum of
the three invariant one-dimensional subspaces corresponding to $J_z=1
,0,-1$.  In this case, three different orbits exist in the
representation, each of them consisting of only one state up to
phases.  However, only the states with $|J_z|=1$ are extremal, whereas
the state with $J_z=0$ is not: as one can easily verify from the fact
that the reduced state is now just the expectation value of $J_z$, an
equal mixture of a $J_z=1$ and a $J_z= -1$ state has the same reduced
state as a $J_z=0$ state, so the latter remains, as in the irreducible
case, generalized entangled.

A generalization of the latter example, which is relevant to the LMG
model we will study in Section \ref{lmg}, is the case of a spin-$J$
system with a distinguished Abelian subalgebra generated by $J_z$ .
Again, one can see that only the states with maximal magnitude of
$J_z$ are generalized unentangled, and only they have maximal purity.

By definition, note that the relative purity and the invariant uncertainty 
functionals as defined in the previous section relate to each other via
\begin{equation}
(\Delta F)^2 = \langle {\cal C}_2 \rangle - P_{\fh} \;,
\label{iu}
\end{equation}
where ${\cal C}_2$ denotes the quadratic Casimir invariant of the Lie 
algebra and $P_\fh$ is given by Eq. (\ref{purity0}) (prior to
rescaling). Because, by standard representation theory, ${\cal
C}_2=c_\ell \openone$,  with $c_\ell \in  {\rlxs}$ within each irrep,
relative purity and invariant uncertainty essentially provide the same
information if $\fh$ acts  irreducibly.  This, however,  is no longer
true in general in the reducible case. In the above  two-spin-1/2
example, for instance, the two measures agree on the singlet  sector;
for triplet states, $J(J+1)=2$, thus the invariant uncertainty value is
1 (same as $P_\fh$) for $|J_z|=1$ (generalized unentangled) states,
whereas it yields 2 for the (zero-purity) state with $J_z=0$ in the 
triplet sector.

\subsection{Extension to mixed states}

For mixed states on ${\cal H}$, the direct generalization of the
squared length of the projection onto $\fh$ as in Eq. (\ref{p0}) does
{\em not} give a GE measure with well-defined monotonicity properties
under appropriate generalizations of the LOCC semigroup of
transformations \cite{baknorvi}. A proper extension of the quadratic
purity measure defined in the previous section for pure states to
mixed states may be naturally obtained via a standard convex roof
construction.  If $\rho = \sum\limits_s p_s \ket{\psi_s}\bra{\psi_s}$,
with $\sum\limits_s p_s =1$ and $\sum\limits_s p_s^2 < 1$, the latter
is obtained by calculating the maximum $\fh$-purity (minimum
entanglement) over all possible convex decompositions $\{p_s,
\ket{\psi_s} \}$ of the density operator $\rho$ as a pure-state
ensemble. In general, similarly to what happens for most mixed-state
entanglement measures, the required extremization makes the resulting
quantity very hard to compute.

While a more expanded discussion of mixed-state GE measures is given
in \cite{baknorvi}, we focus here on applying the notion of GE to
characterize QPTs in different lattice systems.  Because the latter
take place in the limit of zero temperature, the ground state of the
system may be assumed to be pure under ideal conditions.  Accordingly,
Eq. (\ref{purity1}) will suffice for our current purposes.


\section{Relative purity as a measure of entanglement in different 
quantum systems}
\label{examples}

We now apply the concept of relative purity to different physical
systems in order to understand its meaning as a measure of
entanglement for pure quantum states. First, we will concentrate on
spin systems, showing that for particular subsets of observables, the
$\fh$-purity can be reduced to the usual notion of entanglement: the
pure quantum states that can be written as a product of states of each
party will be generalized unentangled.  However, for other physically
natural choices of observable sets, this is no longer the case.  Next,
we study the $\fh$-purity as a measure of entanglement for fermionic
systems, since this is a good starting point for the analysis of the
QPT present in the anisotropic XY model in a transverse magnetic field
(Section \ref{xymodel}). In particular, we show that if a fermionic
state can be represented as a single Slater determinant, it is
generalized unentangled relative to the Lie algebra $\fu (N)$, which
is built from bilinear products of fermionic operators.  These
examples illustrate how the concept and measure of GE is applicable to
systems described by different operator languages, in preparation for
the study of QPTs.

Let us introduce the following representative quantum states for $N$
spins  of magnitude $S$:
\begin{eqnarray}
\label{statedef}
\ket{{\sf F}_S^N}&=&\ket{S,S, \cdots ,S} \;,\\ \nonumber
\ket{{\sf W}_S^N} &= &\frac{1}{\sqrt{N}} \sum\limits_{i=1}^N \ket{S,\cdots,S ,
(S-1)_i , S, \cdots, S}\; ,\\ \nonumber
\ket{{\sf GHZ}_S^N}& = &\frac{1}{\sqrt{2S+1}} \sum\limits_{l=0}^{2S}
\ket{S-l,S-l,\cdots,S-l} \;,
\end{eqnarray}
where the product state $\ket{S_1,S_2,\cdots,S_N}=\ket{S_1}_1 \otimes
\ket{S_2}_2 \otimes \cdots \otimes \ket{S_N}_N$, and $\ket{S_i}_i$ 
denotes the state of the $i$th party with $z$-component of the 
spin equal to $S_i$ (defining the relevant computational basis for
the $i$th subsystem).

\subsection{Two-spin systems}
\label{spinexamples}

\begin{figure}[hbt]
\begin{center}
\includegraphics[height=15cm]{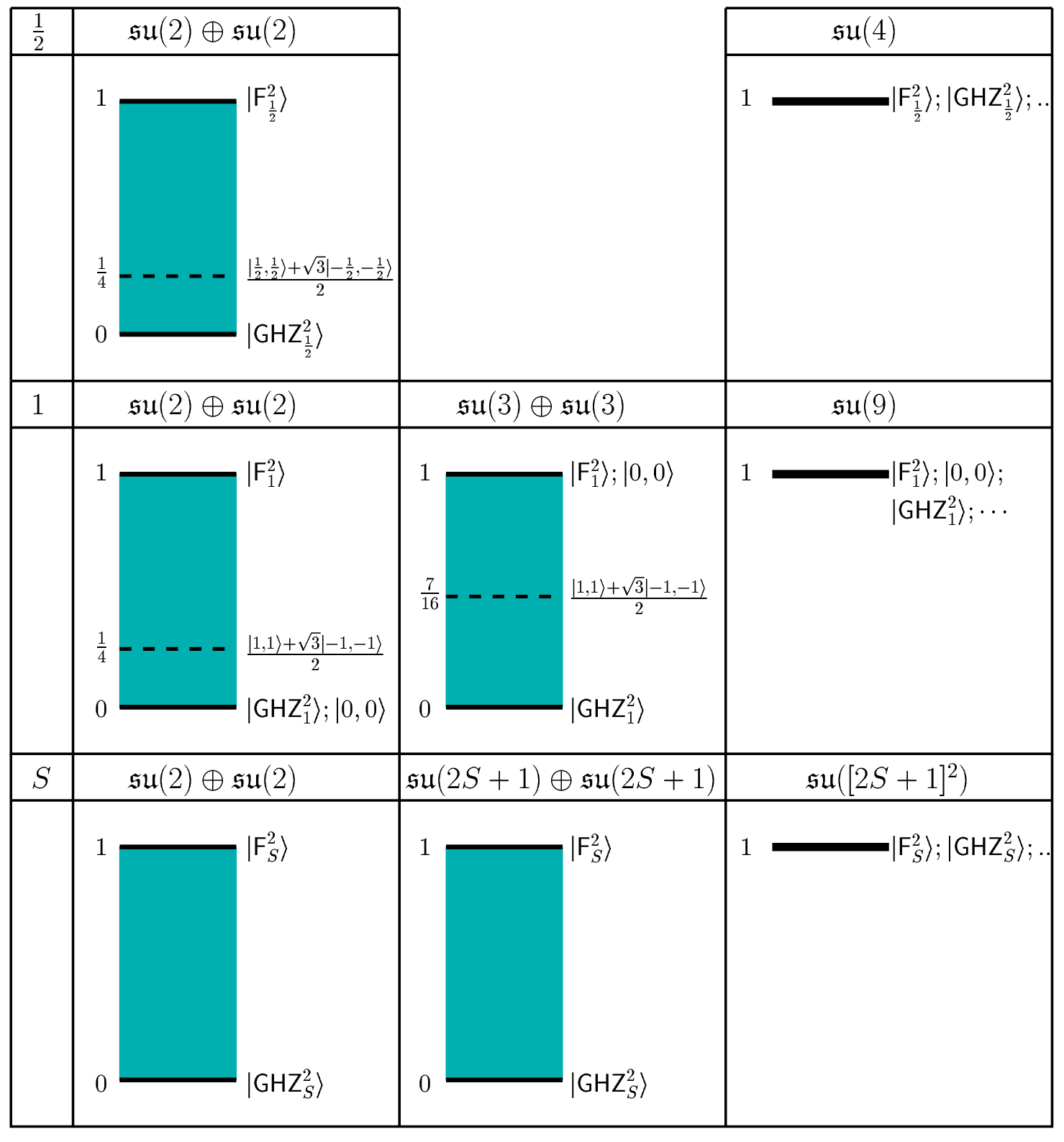}
\end{center}
\caption{Purity relative to different possible algebras for a two-spin-$S$  
system. The quantum states $\ket{{\sf GHZ}_S^2}$ and $\ket{{\sf F}_S^2}$ 
are defined in Eqs. (\ref{statedef}).}
\label{twoparties}
\end{figure}

For simplicity, we begin by studying the GE of a two-qubit system (two
spin-1/2s), where the most general pure quantum state can be written
as  $\ket{\psi}=a \ket{\frac{1}{2},\frac{1}{2}} + b
\ket{\frac{1}{2},-\frac{1}{2}} +c \ket{-\frac{1}{2},\frac{1}{2}} + d
\ket{-\frac{1}{2},-\frac{1}{2}} $, with the complex numbers $a$, $b$,
$c$,  and $d$ satisfying  $|a|^2+|b|^2+|c|^2+|d|^2=1$.  The traditional
measures of  pure-state entanglement in this case are well understood,
indicating that the  Bell states $\ket{\sf GHZ_{\frac{1}{2}}^2}$
\cite{eiporo} (and its local spin  rotations) are maximally entangled
with respect to the local Hilbert space  decomposition ${\cal H}_1
\otimes{\cal H}_2$.  On the other hand, calculating the purity
relative to the (irreducible) Lie algebra of all {\em local 
observables} $\fh=\fsu(2)_1\oplus \fsu(2)_2 =\{\sigma_{\alpha}^i;
\mbox{ } i:1,2;\mbox{ } \alpha=x,y,z\}$ classifies the pure two-spin-1/2  
states in the same way as the traditional measures do (see Fig.
\ref{twoparties}). Here, the operators $\sigma_{\alpha}^1=\sigma_\alpha
\otimes \one $ and $\sigma_\alpha^2=\one \otimes \sigma_\alpha$ are the
Pauli operators acting on spin 1 and 2, respectively, and
\begin{equation}
\label{pauli1}
\one=
\pmatrix {1 & 0 \cr 0 & 1 \cr}, \mbox{ } 
\sigma_x=
\pmatrix {0 & 1 \cr 1 & 0 \cr}, \mbox{ } \sigma_y=
\pmatrix {0 & -i \cr i & 0 \cr},\mbox{ } \sigma_z=
\pmatrix {1 & 0 \cr 0 & -1 \cr}
\;,
\end{equation}
in the basis where $\ket{+1/2}=\ket{\uparrow}=\pmatrix {1 \cr 0}$ and 
$\ket{-1/2}=\ket{\downarrow}=\pmatrix {0 \cr 1}$.  In this case, Eq.
(\ref{purity1})  simply gives 
\begin{equation}
P_{\fh} (\ket{\psi})=\frac{1}{2}\sum\limits_{i,\alpha} \langle 
\sigma_{\alpha}^i \rangle^2 \;, 
\end{equation} 
where Bell's states are maximally entangled ($P_{\fh}=0$) and product
states of  the form $\ket{\psi}=\ket{\phi_1}_1 \otimes \ket{\phi_2}_2$
(GCSs of the local  algebra $\fh$ above) are generalized unentangled, 
with maximum purity.  Therefore, the normalization factor ${\sf K} =
1/2$ may be obtained by setting $P_{\fh}=1$ in such a product state. 
As explained in Section \ref{sectionpurity}, $P_{\fh}$ is invariant
under group operations, i.e., in this case, local rotations. Since
all GCSs of $\fh$ belong to the same orbit generated by the application
of group operations to a particular product state (a reference state
like $\ket{\frac{1}{2},\frac{1}{2}}= \ket{\uparrow,\uparrow}$), 
they all consistently have maximum $\fh$-purity ($P_{\fh}=1$). 

Another important insight may be gained by calculating the purity
relative to the algebra of {\em all} observables for the system,
$\fh=\fsu(4)=\{\sigma_{\alpha}^i,\sigma_{\alpha}^1\otimes
\sigma_{\beta}^2; \mbox{ }i=1,2; \mbox{ }\alpha,\beta=x,y,z\}$ in this
case. One finds that {\em any} two spin-1/2 pure state $\ket{\psi}$
(including Bell's states) is then generalized {\em un}entangled
($P_{\fh}=1$, see also Fig.  \ref{twoparties}).  This property is a
manifestation of the relative nature of GE, as considering the set of
all observables as being physically accessible is equivalent to not
making any preferred subsystem decomposition. Accordingly, in this
case any pure quantum state becomes a GCS of $\fsu(4)$.

In Fig. \ref{twoparties} we also show the GE for systems of two parties
of spin-$S$ relative to different algebras.  We observe that the purity
reduces again to the traditional concept of entanglement for higher
spin if it is calculated relative to the (irreducible) Lie algebra of
{\em all} local observables $\fh=\fsu(2S+1)_1\oplus \fsu(2S+1)_2$. For
example, if we are  interested in distinguishing product states from 
entangled states in a two-spin-1s system, we need to calculate  the
purity relative to the (irreducible) algebra  $\fh=\fsu(3)_1 \oplus
\fsu(3)_2 = \{ \lambda^1_\alpha \otimes \one^2 , \one^1  \otimes
\lambda^2_\alpha \mbox{ } (1 \leq \alpha \leq 8) \}$, where the $3
\times 3$ Hermitian and traceless matrices $\lambda_i$ are the well
known Gell-Mann matrices \cite{georgi}:
\[
\begin{array}{l}
\lambda_1= \frac{1}{\sqrt{2}} 
\pmatrix {0 & 1& 0 \cr 1 & 0 & 0 \cr 0 & 0 & 0 \cr} \mbox{ ; }\;
\lambda_2= \frac{1}{\sqrt{2}} 
\pmatrix {0 & -i & 0 \cr i & 0 & 0 \cr 0 & 0 & 0 \cr} \\
\lambda_3= \frac{1}{\sqrt{2}} 
\pmatrix {1 & 0& 0 \cr 0 & -1 & 0 \cr 0 & 0 & 0 \cr} \mbox{ ; }\;
\lambda_4= \frac{1}{\sqrt{2}} 
\pmatrix {0 & 0& 1 \cr 0 & 0 & 0 \cr 1 & 0 & 0 \cr} \\
\lambda_5= \frac{1}{\sqrt{2}} 
\pmatrix {0 & 0& -i \cr 0 & 0 & 0 \cr i & 0 & 0 \cr} \mbox{ ; }\;
\lambda_6= \frac{1}{\sqrt{2}} 
\pmatrix {0 & 0& 0 \cr 0 & 0 & 1 \cr 0 & 1 & 0 \cr} \\
\lambda_7= \frac{1}{\sqrt{2}} 
\pmatrix {0 & 0& 0 \cr 0 & 0 & -i \cr 0 & i & 0 \cr} \mbox{ ; }\;
\lambda_8= \frac{1}{\sqrt{6}} 
\pmatrix {1 & 0& 0 \cr 0 & 1 & 0 \cr 0 & 0 & -2 \cr} \;,
\end{array}
\]
which satisfy 
${\sf Tr}[ \lambda_\alpha \lambda_\beta ] = \delta_{\alpha,\beta}$.  
In this basis, the computational spin-1 states are represented by the 
3-dimensional vectors
\begin{equation}
\ket{1} = \pmatrix {1 \cr 0 \cr 0 \cr} \mbox{ ; } 
\ket{0} = \pmatrix {0 \cr 1 \cr 0 \cr} \mbox{ and } 
\ket{-1} = \pmatrix {0 \cr 0 \cr 1 \cr}\; .
\end{equation}
Then, the relative purity for a generic pure state $\ket{\psi}$ becomes
\begin{equation}
\label{purityspin1}
P_{\fh} (\ket{\psi})= \frac{3}{4} \sum \limits_{\alpha=1}^8 
\sum \limits_{i=1}^2 \langle \lambda^i_\alpha \rangle ^2\;,
\end{equation}
where $\langle \lambda^i_\alpha \rangle$ denotes the expectation  value
of $\lambda^i_\alpha$ in the state $\ket{\psi}$. In this way, product
states like $\ket{\psi}=\ket{\phi_1}_1 \otimes \ket{\phi_2}_2$ are
generalized unentangled ($P_{\fh}=1$) and states like $\ket{{\sf
GHZ}_1^2}$ (and states connected through local spin unitary
operations),  are maximally entangled in this algebra ($P_{\fh}=0$).

Different results are obtained if the purity is calculated relative to
a {\em subalgebra of local observables}. For example, the two-spin-1
product  state $\ket{0,0}= \ket{0}\otimes\ket{0}$ where both spins have
zero projection along $z$ becomes generalized entangled relative to the
(irreducible) algebra $\fsu(2)_1 \oplus \fsu(2)_2$ of local spin
rotations, which is  generated by $\{ S_{\alpha}^i;\mbox{ }
i:1,2;\mbox{ } \alpha=x,y,z\}$, the spin-1 angular momentum operators
$S_\alpha$ for each spin being given by
\begin{eqnarray}
\label{pauli2}
S_x= \frac{1}{\sqrt{2}}
\pmatrix {0 & 1 & 0 \cr 1 & 0 & 1\cr 0 & 1 & 0\cr}, \mbox{ } 
S_y=\frac{1}{\sqrt{2}}
\pmatrix {0 & -i & 0 \cr i & 0 & -i \cr 0 & i & 0 \cr}, \mbox{ } 
S_z=
\pmatrix {1 & 0 & 0\cr 0  & 0 & 0\cr 0 & 0 & -1 \cr} \;. 
\end{eqnarray}
Notice that access to local angular momentum observables suffices to
operationally characterize the system as describable in terms of two
spin-1 particles (by imagining, for instance, performing a
Stern-Gerlach-type of experiment on each particle). Thus, even when a
subsystem decomposition can be naturally identified from the beginning
in this case, states which are manifestly separable (unentangled) in the 
standard sense may exhibit GE (see also Appendix \ref{app1}). On the 
other hand, this is physically quite natural in the example, since 
there are no SU(2) $\times$ SU(2) group operations (local rotations) 
that are able to transform the state $\ket{0,0}$ into the unentangled 
product state $\ket{1,1}$.

The examples described in this section together with other examples of
states of bipartite quantum systems are shown in
Fig. \ref{twoparties}.  It is clear that calculating the purity
relative to different algebras gives information about different types
of quantum correlations present in the system.


\subsection{$N$-spin systems}
\label{Nspinexamples}

The traditional concept of pure multipartite entanglement in an $N$
spin-$S$ quantum system refers to quantum states that cannot be
written as a product of states of each party. The $\fh$-purity
distinguishes pure product states from entangled ones if it is
calculated relative to the (irreducible) algebra of local observables
$\fh=\bigoplus \limits_{i=1}^N \fsu (2S+1)_i$ (see Appendix \ref{app1}).
By Eq. (\ref{pinvariance}), the measure $P_{\fh}$ is invariant under 
local unitary operations as desired. In particular, the usual concept 
of entanglement in an $N$-qubit quantum state ($N$ spin 1/2s) can be 
recovered if the purity is calculated relative to the local algebra 
$\fh=\bigoplus\limits_{i=1}^N \fsu(2)_i = \{ \sigma_x^1 , \sigma_y^1,
\sigma_z^1, \cdots , \sigma_x^N , \sigma_y^N , \sigma_z^N \}$, where
the Pauli operators $\sigma_\alpha^i$ ($\alpha=x,y,z$) are now
\begin{equation}
\label{pauli3}
\sigma^i_\alpha = \overbrace{\one \otimes \one \otimes \cdots \otimes
\underbrace{\sigma_\alpha}_{i^{th}\ \mbox{factor}} \otimes \cdots
\otimes \one}^{N\ \mbox{factors}} \; ,
\end{equation}
and the $2\times 2$ matrices $\sigma_\alpha$ and $\one$ are given in Eq.
(\ref{pauli1}). Then, the local purity becomes
\begin{equation} 
\label{purity3}
P_{\fh} (\ket{\psi}) = \frac{1}{N} \sum\limits_{\alpha=x,y,z}
\sum\limits_{i=1}^N \langle \sigma_\alpha^i \rangle^2 \;,
\end{equation}
where again the normalization factor ${1}/{N}$ is obtained by
setting $P_{\fh}=1$ in any product state like $\ket{\psi} =
\ket{\phi_1}_1 \otimes \ket{\phi_2}_2 \otimes \cdots \otimes
\ket{\phi_N}_N$ (a GCS in this algebra).  With this definition, 
states like $\ket{{\sf GHZ}_{\frac{1}{2}}^N}$, $[(\ket{\uparrow ,
\downarrow}-\ket{\downarrow , \uparrow})/\sqrt{2}]^{\otimes n}$ (with
obvious notations), and the cluster states $\ket{\Phi}_C$ introduced
in Ref. \cite{brra} (see also Appendix \ref{appb}), will be maximally
entangled ($P_{\fh}=0$).

Remarkably, as announced in \cite{baknorsovi}, after some algebraic 
manipulations (see Appendix \ref{app2}), one can prove that 
\begin{equation}
P_{\fh}(\ket{\psi}) = 1-Q (\ket{\psi}) \:, 
\end{equation}
where $Q$ is the (pure-state) measure of {\em global entanglement} for
$N$ spin 1/2s systems originally introduced by Meyer and Wallach in
\cite{mewa}.  A similar relation was independently derived in
\cite{brennen}.  See also~\cite{vbkos} for additional related 
considerations.

\begin{figure}[b]
\begin{center}
\includegraphics[height=15cm,width=12.5cm]{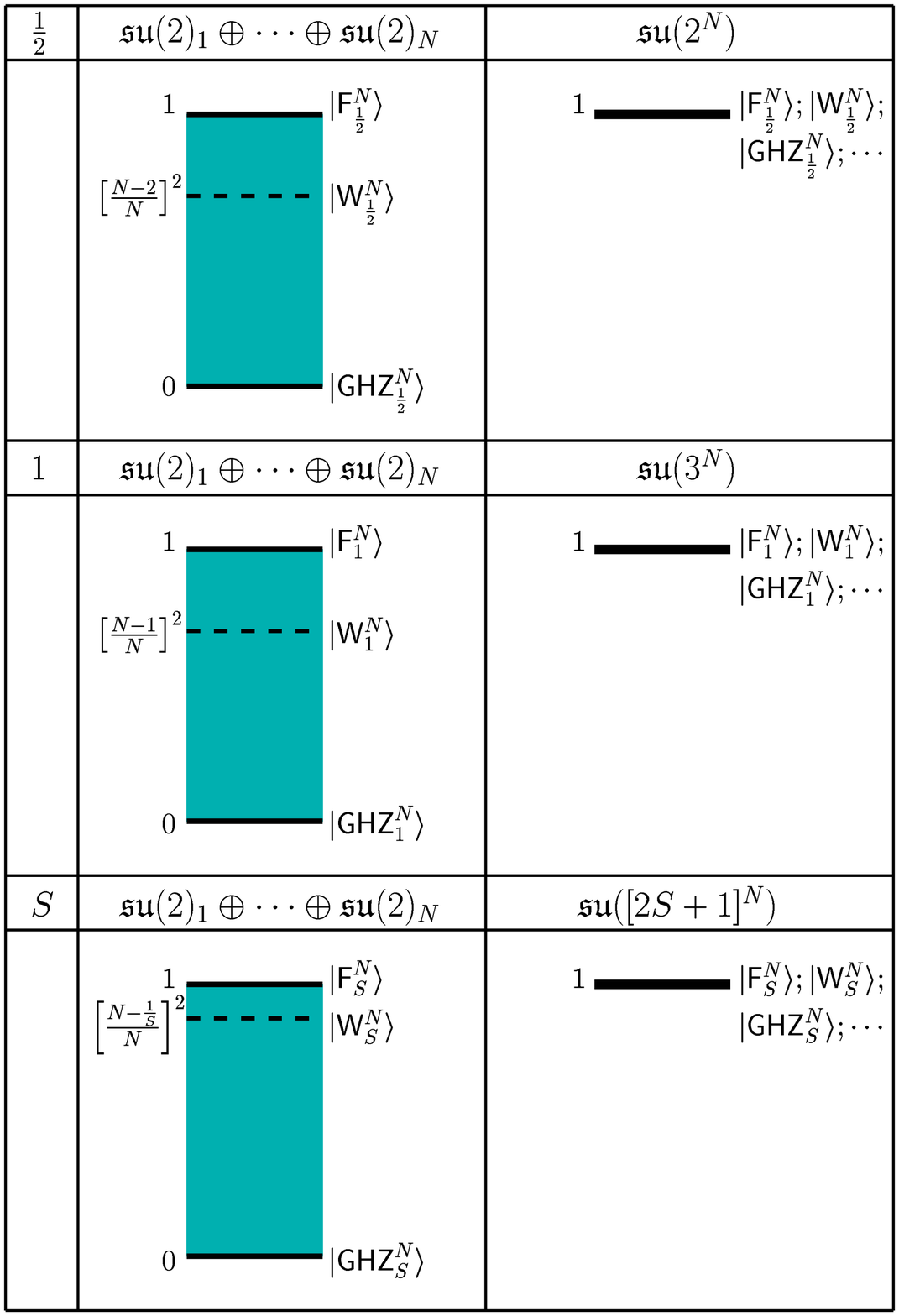}
\end{center}
\caption{Purity relative to different algebras for a $N$ spin-$S$ system.
The quantum states $\ket{{\sf GHZ}_S^N}$, $\ket{{\sf W}_S^N}$, and 
$\ket{{\sf F}_S^N}$ are defined in Eqs. (\ref{statedef}).}
\label{nparties}
\end{figure}

In Fig. \ref{nparties} we display some examples of the purity relative
to the local algebra $\fh=\bigoplus\limits_{i=1}^N \fsu(2)_i$ for a $N$
spin-$S$ system.  We also show the purity relative to the algebra of
all observables $\fsu([2S+1]^N)$,  where any pure quantum state is a
GCS, thus generalized unentangled ($P_{\fh}=1$).

\subsection{Purity relative to the $\fu(N)$ algebra}
\label{sectionunpurity}

We now apply the concept of GE to a physical system consisting of
$N$ (spinless) fermion modes $j$, each mode being described in terms of
canonical creation  and annihilation operators $c^{\dagger}_j$,
$c^{\;}_j$ respectively, satisfying   the following anti-commutation
rules:
\begin{equation} 
\label{anticom}
\{ c^{\dagger}_i , c^{\;}_j \} = \delta_{i,j}\;,\;\;\; \{ c^{\;}_i,
c^{\;}_j \} = 0\;.
\end{equation}
For instance, different modes could be associated with different sites in
a  lattice, or to delocalized momentum modes related to the spatial modes
through a Fourier transform.  In general, for any $N \times N$ unitary
matrix $U$, any transformation $c^{\;}_j \mapsto \sum_j U_{ij}
c^{\;}_j$ maps the original modes into  another possible set of
fermionic modes.  Using the above commutation relations, one also finds
that 
\begin{equation}
[c^{\dagger}_i c^{\;}_j, c^{\dagger}_k c^{\;}_l ] = \delta_{jk}
c^{\dagger}_i c^{\;}_l - \delta_{il} c^{\dagger}_k c^{\;}_j \;.
\end{equation}
Thus, the set of bilinear fermionic operators $\{ c^\dagger_j
c^{\;}_{j'}; \; 1 \leq j,j' \leq N \}$ provides a realization of the
unitary Lie algebra  $\fu(N)$ in the $2^N$-dimensional Fock space
${\cal H}_{Fock}$ of the system.  The latter is constructed as the
direct sum of subspaces ${\cal H}_n$  corresponding to a fixed fermion
number $n=0,\ldots,N$, with dim(${\cal H}_n ) = N!/[n! (N-n)!]$.   For
our purposes, it is convenient to express $\fu(N)$ as the linear span of  
a Hermitian, orthonormal operator basis, which we choose as follows:
\begin{equation}
\fu(N)=\left\{
\begin{array}{cl}
(c^{\dagger}_j c^{\;}_{j'} + c^{\dagger}_{j'} c^{\;}_j) & 
\mbox{  with } 1\leq j<j' \leq N \cr
i(c^{\dagger}_j c^{\;}_{j'} - c^{\dagger}_{j'} c^{\;}_j) & 
\mbox{  with } 1\leq j<j' \leq N \cr 
\sqrt{2}(c^{\dagger}_j c^{\;}_j - 1/2 ) & \mbox{  with }1 \leq j \leq N
\end{array}
\right. \;,
\label{uNbasis}
\end{equation}
(We use henceforth the notational convention that the large left curly 
bracket means ``is the span of'').  The action of $\fu(N)$ on ${\cal
H}_{Fock}$ is reducible, because any operator in $\fu(N)$ conserves
the total number of fermions $n=\langle \sum_{j=1}^N c^\dagger_j c^{\;}_j
\rangle$. It turns out that the irrep decomposition of $\fu(N)$ is
identical to the direct sum into fixed-particle-number subspaces
${\cal H}_n$, each irrep thus appearing with multiplicity one.

Using Eq. (\ref{purity1}), the $\fh$-purity of a generic pure
many-fermion state relative to $\fu(N)$ becomes 
\begin{equation}
\label{unpurity}
 P_{\fh}(\ket{\psi}) = \frac{2} {N} \sum \limits_{j<j'=1}^N  
\Big[ \langle c^{\dagger}_j c^{\;}_{j'} +
 c^{\dagger}_{j'} c^{\;}_j \rangle^2  -  \langle c^{\dagger}_j c^{\;}_{j'} -
 c^{\dagger}_{j'} c^{\;}_j \rangle^2 \Big] +\frac{4}{N} 
\sum\limits_{j=1}^N \langle
 c^{\dagger}_j c^{\;}_j -1/2 \rangle^2\;.
\end{equation}
Here, we took ${\sf K}= 2/N$, for reasons that will become clear
shortly. In this case, the fermionic product states (Slater
determinants) of the form $\ket{\phi}= \prod\limits_l c^\dagger_l
\ket{\sf vac}$, with $\ket{\sf vac}$ denoting the reference state 
with no fermions and $l$ labelling a particular set of modes, are 
the GCSs of the $\fu(N)$ algebra \cite{gi,pe}. Because a Slater
determinant carries a well defined number of particles, each GCS
belongs to an irrep space ${\cal H}_n$ for some $n$, states with
different $n$ belonging to different orbits under $\fu (N)$.  A fixed
GCS has maximum $\fh$-purity when compared to any other state
within the same irrep space.  Remarkably, it also turns out that
any GCS of $\fh=\fu(N)$ gives rise to a reduced state which is
extremal (thus generalized unentangled) regardless of $n$, the
$\fh$-purity assuming the same (maximum) value in each irrep. Using
this property, the normalization factor ${\sf K}={2}/{N}$ was
calculated by setting $P_{\fh}=1$ in an arbitrary Slater determinant.
Thus, the purity relative to the $\fu(N)$ algebra is a good measure of
entanglement in fermionic systems, in the sense that $P_{\fh}=1$ in
any fermionic product state, and $P_{\fh}<1$ for any other state,
irrespective of whether the latter has a well defined number of fermions
or not.  Notice that, thanks to the invariance of $P_{\fh}$ under group
transformations (Eq. (\ref{pinvariance})), the property of a state
being generalized unentangled is independent of the specific set of
modes that is chosen.  This is an important difference between our GE
and the mode entanglement approach~\cite{zanardi,zanardi2}.

\section{Entanglement and quantum phase transitions}
\label{purity-QPT}

As already mentioned, although many measures of entanglement have been
defined in the literature, assessing their ability to help us better
understand QPTs in quantum systems largely remains an open problem. In
the following two sections we attempt to characterize the QPTs present
in the LMG model and in the anisotropic XY model in an external
magnetic field through the GE notion, relative to a particular subset
of observables which will be appropriately chosen in each case.
Interestingly, for both these models the ground states can be computed
exactly by mapping the set of observable operators involved in the
system Hamiltonian to a new set of operators which satisfy the same
commutation relations, thus preserving the underlying algebraic
structure. In the new operator language, the models are seen to
contain some symmetries that make them exactly solvable, allowing one
to obtain the ground state properties in a number of operations that
scales polynomially with the system size (see also \cite{somma} for
related discussions).  It is possible then to understand which quantum
correlations give rise to the QPTs in these cases.

Several issues should be considered when looking for an algebra $\fh$
of observables that may make the corresponding relative purity a good
indicator of a QPT.  A first relevant observation is that in each of
these cases a preferred Lie algebra exists, where the respective
ground state would have maximum $\fh$-purity independently of the
interaction strengths in the Hamiltonian. The purity relative to such
an algebra remains constant, therefore it does not identify the
QPT.  (In these cases, this algebra is in fact the Lie algebra generated
by the parametrized family of model Hamiltonians, as the parameters are
varied.)  Thus, one needs to extract a subalgebra relative to which the
ground state may be generalized entangled, depending on the parameters
in the Hamiltonian.  A second, closely related observation is that the
purity must contain information about quantum correlations which
undergo a qualitative change as the transition point is crossed: thus,
the corresponding degree of entanglement, as measured by the purity,
must depend on the interaction strengths governing the phase
transition.  Finally, whenever a degeneracy of the ground state exists
or emerges in the thermodynamic limit, a physical requirement is that 
the purity be the same for all ground states.

Although these restrictions together turn out to be sufficient for
choosing the relevant algebra of observables in the following two
models, they do not provide an unambiguous answer when solving a
non-integrable model whose exact ground state solution cannot be
computed efficiently.  Typically, in the latter cases the ground
states are GCSs of Lie algebras each of whose dimension increases
exponentially with the system size.  Choosing the observable
subalgebra that contains the proper information on the QPTs 
(such as information on critical exponents) then becomes, 
in general, a difficult task.

On the other hand, a concept of {\em generalized mean-field Hamiltonian}
emerges from these considerations. Given a Hilbert space ${\cal H}$ of
dimension $p^N$ (with $p$ an integer $>1$), we will define a mean-field
Hamiltonian as an operator
\begin{equation}
H_{\sf MF}=\sum_\alpha \epsilon_\alpha A_\alpha \: , \hspace{5mm}
\epsilon_\alpha \in \mathbb{R} \:,
\end{equation}
that is an element of an irreducibly represented Lie algebra of 
Hermitian operators
$\fh=\{A_1,\cdots,A_L\}$ whose dimension scales polynomially in $N$
that is, $L=\mbox{poly}(N)$. When the ground state of such an
$H_{\sf MF}$ is non-degenerate, it turns out to be a GCS 
of $\fh$ \cite{baknorvi}, while the
remaining eigenstates 
(some of which may also be GCSs) and energies can be efficiently
computed. The connection between Lie-algebraic mean-field Hamiltonians
and their efficient solvability deserves a careful analysis in its
own right, which we will present elsewhere~\cite{next}.

\section{Lipkin-Meshkov-Glick model}
\label{lmg}

Originally introduced in the context of nuclear physics \cite{limegl},
the Lipkin-Meshov-Glick (LMG) model is widely used as a testbed for
studying critical phenomena in (pseudo)spin systems \cite{gi}. This model
was shown to be exactly-solvable in \cite{orsoduro}. In this
section, we investigate the critical properties of this model by
calculating the purity relative to a particular subset of observables,
which will be chosen by analyzing the {\em classical} behavior of the ground
state of the system. For this purpose, we first need to map the model
to a {\em single} spin, where it becomes solvable and where the
standard notion of entanglement is not immediately applicable.

The model is constructed by considering $N$ fermions distributed in two
$N$-fold  degenerate levels (termed upper and lower shells).  The
latter are separated by  an energy gap $\epsilon$, which will be set
here equal to 1.  The quantum number  $\sigma =\pm 1$ ($\uparrow$ or
$\downarrow$) labels the level while the quantum  number $k$ denotes
the particular degenerate state in the level (for both shells,  $k \in
\{k_1,\ldots,k_N\}$).  In addition, we consider a ``monopole-monopole''
interaction that scatters pairs of particles between the two levels
without  changing $k$. The model Hamiltonian may  be written as 
\begin{equation}
\label{lmghamilt1}
H=H_0 + \hat{V} + \hat{W} =
\frac{1}{2} \sum\limits_{k,\sigma} \sigma c^\dagger_{k \sigma}c^{\;}_{k \sigma}
+ \frac{V}{2N} \sum\limits_{k,k',\sigma} c^\dagger_{k \sigma}
c^\dagger_{k' \sigma} c^{\;}_{k'\overline{\sigma} } 
c^{\;}_{k \overline{\sigma}} + 
\frac{W}{2N}
\sum\limits_{k,k',\sigma} c^\dagger_{k \sigma}
c^\dagger_{k' \overline{\sigma}} c^{\;}_{k' \sigma} 
c^{\;}_{k \overline{\sigma}} \;, 
\end{equation}
where $\overline{\sigma} = -\sigma$, and the fermionic operators 
$c^\dagger_{k \sigma}$ ($c^{\;}_{k \sigma}$) create (annihilate) a
fermion in the  level identified by the quantum numbers $(k,\sigma)$
and satisfy the fermionic  commutation relations given in Section
\ref{sectionunpurity}.  Thus, the  interaction $\hat{V}$ scatters a
pair of particles belonging to one of the  levels, and the interaction
$\hat{W}$ scatters a pair of particles belonging to   different levels.
Note that the factor $1/N$ must be present in the interaction  terms
for stability reasons, as the energy per particle must be finite in 
the thermodynamic limit.

Upon introducing the pseudospin operators
\begin{eqnarray}
\label{pseudospin1}
J_+ &=& \sum\limits_k c^\dagger_{k \uparrow} c^{\;}_{k \downarrow}\;, \\
\label{pseudospin2}
J_- &=& \sum\limits_k c^\dagger_{k \downarrow} c^{\;}_{k \uparrow}\;, \\
\label{pseudospin3}
J_z &=& \frac{1}{2} \sum\limits_{k,\sigma} \sigma c^\dagger_{k \sigma} 
c^{\;}_{k \sigma}
= \frac{1}{2} \Big( n_\uparrow -n_\downarrow \Big) \;,
\end{eqnarray}
which satisfy the $\fsu(2)$ commutation relations of the angular momentum 
algebra, 
\begin{eqnarray}
\left[ J_z, J_{\pm} \right] &=& \pm J_{\pm}\;, \label{angmom0}\\
\left[ J_+ , J_- \right] &=& J_z \;,
\label{angmom}
\end{eqnarray}
the Hamiltonian of Eq. (\ref{lmghamilt1}) may be rewritten as
\begin{equation}
\label{lmghamilt2}
H=J_z + \frac{V}{2N} (J_+^2 + J_-^2 ) + \frac{W}{2N} (J_+ J_- + J_- J_+)\;.
\end{equation}
As defined by Eq. (\ref{lmghamilt2}), $H$ is invariant under the
$\zlxs_2$ inversion symmetry operation $K$ that transforms
$(J_x,J_y,J_z) \mapsto (-J_x,-J_y,J_z)$, and it also commutes with the
(Casimir) total angular momentum operator ${\bf J}^2 = J_x^2
+J_y^2+J_z^2$. Therefore, the non-degenerate eigenstates of $H$ are
simultaneous eigenstates of both $K$ and ${\bf J}^2$, and they may be
obtained by diagonalizing matrices of dimension $2J+1$ (whereby the
solubility of the model).  Notice that, by definition of $J_z$ as in
Eq.  (\ref{pseudospin3}), the maximum eigenvalue of $J_z$ and $J=|{\bf
J}|$ is $N/2$.  In particular, for a system with $N$ fermions as
assumed, both the ground state $\ket{g}$ and first excited state
$\ket{e}$ belong to the largest possible angular momentum eigenvalue
$J=N/2$ \cite{limegl} (so-called half-filling configurations); thus,
they can be computed by diagonalizing a matrix of dimension $N+1$.

\begin{figure}[t]
\begin{center}
\includegraphics[height=5cm]{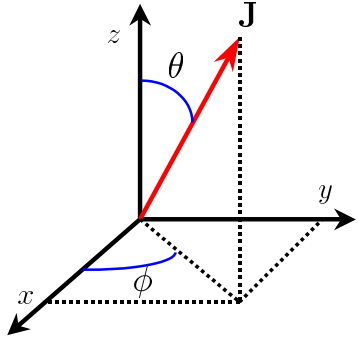}
\end{center}
\caption{Angular momentum coordinates in the three-dimensional space.}
\label{angmomcoord}
\end{figure}

\begin{figure}[hbt]
\begin{center}
\includegraphics[width=12cm]{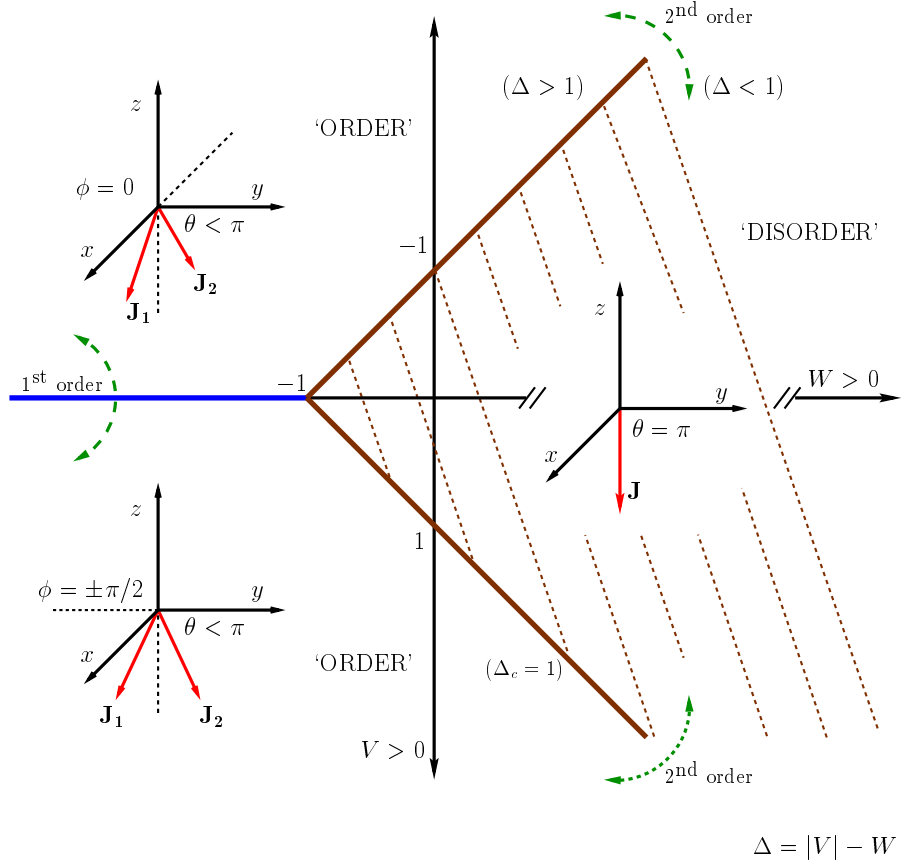}
\end{center}
\caption{Representation of the classical ground state of the LMG model.}
\label{classicreplmg}
\end{figure}

The Hamiltonian (\ref{lmghamilt2}) does not exhibit a QPT for finite
$N$. It is  important to remark that some critical properties of the
LMG model in the thermodynamic limit $N\rightarrow \infty$ can be
understood by using a semiclassical approach \cite{gi2} (note that 
the critical behavior is essentially mean-field): first, we
replace the angular momentum  operators in $H/N$ (with $H$ given in
Eq.(\ref{lmghamilt2})) by their classical  components (Fig.
\ref{angmomcoord}); that is
\begin{eqnarray}
\label{angmomtrans}
{\bf J}=(J_x,J_y,J_z) &\rightarrow &\left(  J\sin \theta \cos \phi , J \sin
\theta \sin \phi, J \cos \theta \right) \;, \\
H/N &\rightarrow & h_c(j,\theta,\phi)\;, 
\end{eqnarray}
where $h_c$ is the resulting classical Hamiltonian and $j=J/N$,
$j=0,\ldots,1/2$.   In this way, one can show that in the thermodynamic
limit (see Appendix \ref{app3}) 
\begin{equation}
\label{lmgenergy}
\lim_{N \rightarrow \infty} \frac{\langle g | H | g  \rangle}{N}  =
\lim_{N \rightarrow \infty} \frac{E_g}{N} = \min_{j ,\theta,\phi}
h_c(j,\theta,\phi)\;,
\end{equation}
so the ground state energy per particle $E_g/N$ can be easily evaluated 
by minimizing 
\begin{equation}
\label{classichamilt}
h_c (j, \theta, \phi) = j \cos \theta  + \frac{V}{2}j^2 \sin^2 \theta  \cos(2 \phi)
+Wj^2 \sin^2  \theta \;.
\end{equation} 
As mentioned, the ground and first excited states have maximum angular
momentum  $j=1/2$. In Fig. \ref{classicreplmg} we show the orientation
of the angular  momentum in the ground states of the classical
Hamiltonian $h_c$,  represented by the vectors ${\bf J}$, ${\bf J_1}$,
and  ${\bf J_2}$, for different values of $V$ and $W$. When $\Delta
=|V|-W \leq 1$ we have $\theta = \pi$ and  the classical angular
momentum is oriented in the negative $z$-direction. However, when
$\Delta > 1$ we have $\cos \theta  = -\Delta^{-1}$  and the classical
ground state becomes two-fold degenerate (notice that $h_c$ is
invariant under the  transformation $\phi \mapsto - \phi$). In this
region and for $V<0$ the angular momentum is oriented in the $xz$
plane  ($\phi=0$) while for $V>0$ it is oriented in the $yz$ plane
($\phi = \pm\pi/2$). The model has a gauge symmetry in the line $V=0$,
$W<-1$, where $\phi$ can take any possible value.

\subsection{First and second order QPTs, and critical behavior}
\label{criticallmg}

Going back to the original Hamiltonian of Eq. (\ref{lmghamilt1}), the
quantum  system undergoes a second order QPT at the critical boundary
$\Delta_c=|V_c|-W_c=1$,  where for $\Delta>\Delta_c$ the ground and
first excited states $\ket{g}$ and $\ket{e}$ become degenerate in the
thermodynamic limit and the inversion symmetry $K$ breaks.   The order
parameter is given by the mean number of fermions in the upper shell 
$\langle n_\uparrow \rangle = 1/2+\langle J_z \rangle /N$, which  in
the thermodynamic limit converges to its classical value, 
\begin{equation}
\label{lmgop}
\lim_{N\rightarrow \infty} 
\langle n_\uparrow \rangle= \frac{1+\cos\theta}{2}\;.
\end{equation}
Obviously, for $\Delta \leq \Delta_c$ we have $\langle n_\uparrow
\rangle=0$, and  $\langle n_\uparrow \rangle > 0$ otherwise 
(see Fig.\ref{classicreplmg}). The critical exponents of the order parameter
are easily computed by making a Taylor expansion near the  critical
points ($\Delta \rightarrow 1^+$). Defining the quantities $x=V_c-V$ 
and $y=W_c-W$, we obtain
\[
\lim_{\Delta \rightarrow 1^+} \langle n_\uparrow \rangle =
\left \{
\begin{array}{l}
(y^\alpha-x^\beta)/2 \hspace{3mm}\mbox{ for }V>0\\ 
(y^\alpha+x^\beta)/2 \hspace{3mm}\mbox{ for }V<0 \end{array}
\right. ,
\]
where the critical exponents are $\alpha=1$ and $\beta=1$.

\begin{figure}[hbt]
\begin{center}
\includegraphics[width=13.8cm]{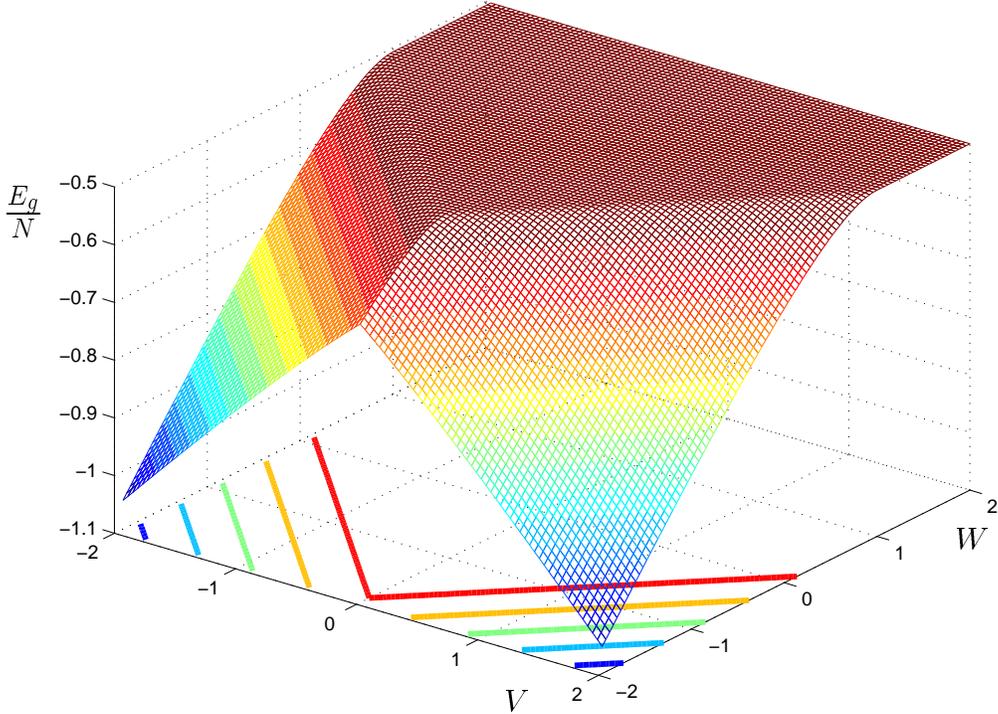}
\end{center}
\caption{Ground state energy per particle in the LMG model.}
\label{lmgenergyplot}
\end{figure}

In Fig. \ref{lmgenergyplot} we show the exact ground state energy per
particle $E_g/N$ (with $E_g = \langle g | H | g\rangle$) as a function
of $V$ and $W$ in the  thermodynamic limit (Eqs. (\ref{lmgenergy})).
One can see that also in the broken symmetry region ($\Delta>1$)
the system undergoes a first order QPT at $V=0$; that is, the first
derivative of the ground state energy with respect to $V$ is not
continuous in this line.

\subsection{Purity as an indicator of the QPTs in the LMG model}
\label{jzpurity}

The standard notion of entanglement is not directly applicable to the
LMG model as described by Eq. (\ref{lmghamilt2}), for this is a single
spin system and no physically natural partition into subsystems is
possible.   Therefore, using the $\fh$-purity as a measure of
entanglement becomes an advantage from this point of view, since the
latter only depends on a particular subset of observables and no
partition of the system is necessary.  The first required step is the
identification of a relevant Lie algebra of observables relative to
which the purity has to be calculated.

Since both the ground and first excited states of the quantum LMG model
may be understood as states of a system carrying total angular
momentum $J=N/2$, a  first possible algebra to consider is the
$\fsu(N+1)$ algebra acting on the  relevant $(N+1)$-dimensional
eigenspace.  Relative to this algebra, $|g\rangle$ is generalized
unentangled for arbitrary values of $V,W$ thus the corresponding purity
remains constant and does not signal the QPTs.  However, the 
family of Hamiltonians (\ref{lmghamilt2}) do not generate this 
Lie algebra, but rather an $\fsu(2)$ algebra, so perhaps $\fsu(N+1)$ is
not a natural choice physically~\cite{jvidal}.

Thus a natural choice, suggested by the commutation
relationships of Eqs. (\ref{angmom0}) and (\ref{angmom}), is to study
the purity relative to the spin-$N/2$  representation of the angular
momentum Lie algebra  $\fh=\fsu(2)= \{ J_x,J_y,J_z \}$:
\begin{equation}
P_{\fh} (\ket{\psi}) = \frac{4}{N^2}\Big[ \langle J_x \rangle ^2 
+\langle J_y \rangle ^2 + \langle J_z \rangle ^2 \Big] \;,
\end{equation}
where the normalization factor ${\sf K}=N^2/4$ is chosen to ensure
that the maximum of $P_{\fh}$ is equal to 1.  With this normalization
factor, $P_{\fh}$ can be calculated exactly in the thermodynamic limit
by relying on the semi-classical approach described earlier (see
Appendix \ref{app3} and Eq. (\ref{angmomtrans})).  For $V=0$ and
arbitrary $W>0$, $|g\rangle= | J_z =-N/2\rangle$ which is a GCS of
$\fsu(2)$ and has $P_{\fh}=1$. For generic interaction values such
that $\Delta \leq 1$, the classical angular momentum depicted in Fig.
\ref{classicreplmg} is oriented along the $z$-direction and is not
degenerate: because $\langle J_x \rangle= \langle J_y \rangle= 0$,
only $\langle J_z \rangle$ contributes to $P_{\fh}$; by recalling that
$ \lim_{N \rightarrow \infty}\langle J_z/N \rangle= -1/2$, this gives
$P_{\fh}=1$, so that {\em as far as relative purity is concerned the
ground state behaves asymptotically like a coherent state in the
thermodynamic limit}.  Physically, this means that with respect to the
relevant fluctuations, GCSs of $\fsu(2)$ are a good approximation of
the quantum ground state for large particle numbers, as is well
established for this model \cite{feng}.  However, in the region
$\Delta > 1$ the ground state (both classical and quantum) is two-fold
degenerate in the $N \rightarrow \infty$ limit, and the value of
$P_{\fh}$ depends in general on the particular linear combination of
degenerate states.  This can be understood from
Fig. \ref{classicreplmg}, where different linear combinations of the
two degenerate vectors ${\bf J_1}$ and ${\bf J_2}$ imply different
values of $\langle J_x \rangle $ for $V<0$ and different values of
$\langle J_y \rangle $ for $V>0$, while $\langle J_z \rangle$ remains
constant.  With these features, the purity relative to the $\fsu(2)$
algebra will not be a good indicator of the QPT.

An alternative option is then to look at a subalgebra of $\fsu(2)$. In 
particular, if we only consider the purity relative to the single
observable  $\fh=\fso(2)=\{J_z\}$ (i.e., a particular CSA of
$\fsu(2)$), and retain the  same normalization as above, we have 
\begin{equation}
\label{jzpuritylmg}
P_{\fh}(\ket{\psi}) =\frac{4}{N^2} \langle J_z \rangle ^2 \;,
\end{equation}
This new purity will be a good indicator of the QPT, since $P_{\fh}=1$
only for $\Delta \leq 1$ in the thermodynamic limit, and in addition
$P_{\fh}$ does not  depend on the particular linear combination of the
two-fold degenerate states in  the region $\Delta > 1$, where
$P_{\fh}<1$. Obviously, in this case $P_{\fh}$ is straightforwardly
related to the order parameter (Eq. (\ref{lmgop})); the critical
exponents of $P_{\fh}-1$ are indeed the same ($\alpha=1$ and
$\beta=1$).

Note that the purity defined by Eq. (\ref{jzpuritylmg}) does not
always take its maximum value for GCSs of $\fh=\fso(2)$ (eigenstates
of $J_z$). In the region $\Delta <1 $ where $P_{\fh} =1$, the quantum
ground state of the LMG model (Eq. (\ref{lmghamilt2})) does not have a
well defined $z$-component of angular momentum except at $V= 0$ ($[ H,
J_z ] \neq 0$ if $V \neq 0$), thus in general it does not lie on a
coherent orbit of this algebra for finite $N$.  However, as discussed
above, it behaves asymptotically (in the infinite $N$ limit) as a GCS
(in the sense that $P_{\fh}\rightarrow 1$).  Moreover, in Section
\ref{sectionpurity} we showed that for $J_z$-eigenstates with
eigenvalues $|J_z|<N/2$, we also obtain $P_{\fh}<1$.

\begin{figure}[hbt]
\begin{center}
\includegraphics[width=12cm]{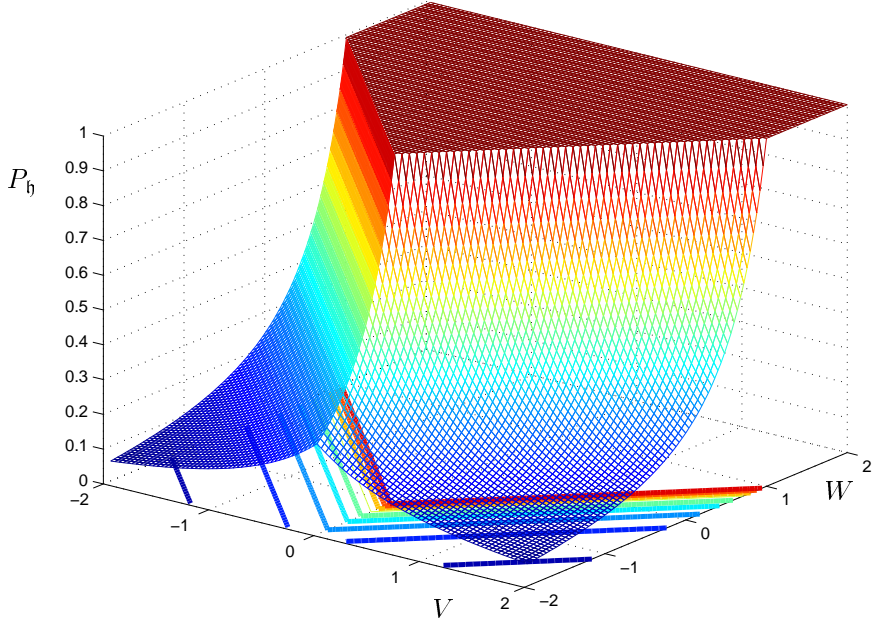}
\end{center}
\caption{Purity relative to the observable $J_z$ in the ground state 
of the LMG model.}
\label{lmgpurityplot}
\end{figure}

In Fig. \ref{lmgpurityplot} we show the behavior of $P_{\fh}$ as a
function of  the parameters $V$ and $W$. Interestingly, the purity
relative to $J_z$ is a good indicator not only of the second order QPT
but also of the first order QPT (the line $V=0$, $W<-1$).  


\vspace*{3mm}

\section{Anisotropic XY model in a transverse magnetic field}
\label{xymodel}

In this section, we exploit the purity relative to the $\fu(N)$
algebra (introduced in Section (\ref{sectionunpurity})) as a measure
able to identify the paramagnetic to ferromagnetic QPT in the
anisotropic one-dimensional spin-1/2 XY model in a transverse magnetic
field and classify its universality properties.

\begin{figure}[t]
\begin{center}
\includegraphics[width=11.5cm]{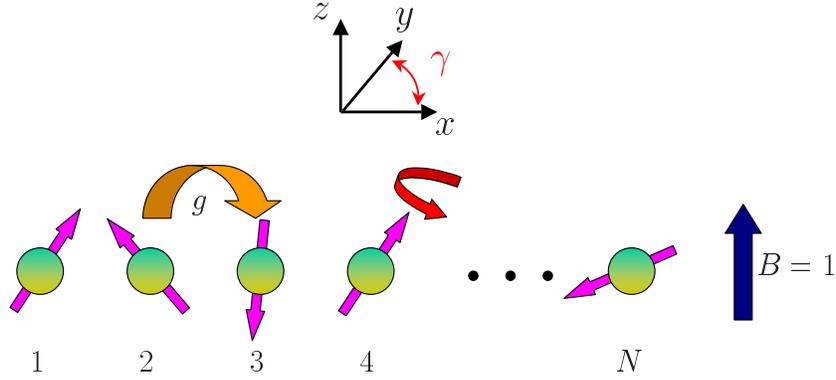}
\end{center}
\caption{Anisotropic one-dimensional XY model in an external transverse
magnetic field $B$.}
\label{figxymodel}
\end{figure}

The model Hamiltonian for a chain of $N$ sites is given by (see
Fig.\ref{figxymodel})
\begin{equation} 
\label{Ham1} 
H =-g \sum\limits_{i=1}^N  \Big[(1+\gamma)  \sigma_x^i \sigma_x^{i+1}+ (1-\gamma) 
\sigma_y^i \sigma_y^{i+1}\Big] + \sum\limits_{i=1}^N \sigma_z^ i \;,
\end{equation}
where the operators  $\sigma_{\alpha}^i$ ($\alpha =x,y,z$)  are the
Pauli spin-1/2 operators on site $i$ (defined in Eqs. (\ref{pauli1})
and (\ref{pauli3})),  $g$ is the parameter one may tune to drive the
QPT, and $0 < \gamma \leq 1 $ is the amount of anisotropy in the
$xy$ plane. In particular, for $\gamma=1$ Eq. (\ref{Ham1}) reduces to
the Ising  model in a transverse magnetic field, while for $\gamma\rightarrow 0$
the model becomes isotropic. Periodic boundary conditions were
considered here, that is  $\sigma_{\alpha}^{i+N} =
\sigma_{\alpha}^{i}$, for all $i$ and $\alpha$.

When $g \gg 1 $ and $\gamma=1$ the model is Ising-like.
In this limit, the spin-spin
interactions are the dominant  contribution to the Hamiltonian
(\ref{Ham1}), and the ground state becomes  degenerate in the
thermodynamic limit, exhibiting ferromagnetic long-range order 
correlations in the $x$ direction:  $M_x^2=\lim_{N\rightarrow \infty}
\langle \sigma_x^1 \sigma_x^{N/2} \rangle > 0$, where $M_x$ is the
magnetization in the $x$-direction. In the opposite limit  where $g
\rightarrow 0$, the external magnetic field becomes important, the
spins  tend to align in the $z$ direction, and the magnetization in the $x$
direction vanishes:  $M_x^2 =\lim_{N\rightarrow \infty} \langle
\sigma_x^1 \sigma_x^{N/2} \rangle = 0$. Thus, in the thermodynamic
limit the model is subject to a paramagnetic-to-ferromagnetic  second
order QPT at a critical point $g_c$ that will be determined later, 
with critical behavior belonging to the 2-$D$ Ising
model universality class. 

This model can be exactly solved using the Jordan-Wigner transformation
\cite{jowi},  which maps the Pauli (spin 1/2) algebra into the
canonical fermion  algebra through 
\begin{equation}  
c^{\dagger}_j=  \prod\limits_{l=1}^{j-1} (-\sigma_z^l)
 \sigma_+^j \;, 
\end{equation} 
where the fermionic operators $c^{\dagger}_j$ ($c^{\;}_j$) have been
introduced in  Section \ref{sectionunpurity} and $\sigma_+^j =
(\sigma_x^j + i \sigma_y^j)/2$ is the raising spin operator. 

In order to find the exact ground state, we first need to write the
Hamiltonian given in Eq. (\ref{Ham1}) in terms of these fermionic
operators, 
\begin{equation}
\label{Ham2}
H = -2g \sum\limits_{i=1}^{N-1} ( c^{\dagger}_i c^{\;}_{i+1}   + \gamma
c^{\dagger}_i c^{\dagger}_{i+1} + h.c.)+ 2g K (c^{\dagger}_N c^{\;}_1 + \gamma
c^{\dagger}_N c^{\dagger}_1 + h.c.) + 2 \hat{N}\;,
\end{equation}
where $K=\prod\limits_{j=1}^N (-\sigma_z^j)$ is an operator that
commutes with the Hamiltonian, and $\hat{N}=\sum\limits_{i=1}^N
c^\dagger_i c^{\;}_i$ is the total number operator (here, we choose
$N$ to be even). Then, the eigenvalue of $K$ is a good quantum number,
and noticing that $K= e^{i \pi \hat{N} }$ we obtain $K=+1 (-1)$
whenever the (non-degenerate) eigenstate of $H$ is a linear
combination of states with an even (odd) number of fermions.  In
particular, the numerical solution of this model in finite systems
(with $N$ even) indicates that the ground state has eigenvalue $K=+1$,
implying anti-periodic boundary conditions in Eq. (\ref{Ham2}).

The second step is to re-write the Hamiltonian in terms of the
fermionic operators $\tilde{c}^{\dagger}_{k}$
($\tilde{c}^{\;}_{k}$), defined by the   Fourier transform of the
operators $c^{\dagger}_j$ ($c^{\;}_j$):
\begin{equation}
\tilde{c}^{\dagger}_{k}= \frac{1} {\sqrt{N}} \sum \limits_{j=1}^N e^{-ikj} 
c^{\dagger}_j \;,
\end{equation}
where the set $V$ of possible $k$ is determined by the anti-periodic
boundary conditions in the fermionic operators:  $V=V_+ + V_-=[ \pm
\frac{\pi}{N}  , \pm \frac {3 \pi}{N}, \cdots, \pm \frac  {(N-1)\pi}{N}
]$. Therefore, we rewrite the Hamiltonian as
\begin{equation}
\label{Ham3}
H +N = -2 \sum\limits_{k \epsilon V} (-1+2g \cos k) \tilde{c}^{\dagger}_k  
\tilde{c}^{\;}_k +  
i g \gamma \sin k (\tilde{c}^{\dagger}_{-k} \tilde{c}^{\dagger}_k 
+  \tilde{c}^{\;}_{-k} \tilde{c}^{\;}_k )\;.
\end{equation}

The third and final step is to diagonalize the Hamiltonian of Eq.
(\ref{Ham3}) using the Bogoliubov canonical transformation
\[ \left \{
\begin{array}{l}
\gamma_k = u_k \tilde{c}^{\;}_k - i v_k \tilde{c}^{\dagger}_{-k} \\
\gamma^{\dagger}_{-k} = u_k \tilde{c}^{\dagger}_{-k} - i v_k \tilde{c}^{\;}_k
\end{array}
\right. \;,
\]
where the real coefficients $u_k$ and $v_k$ satisfy the relations
\begin{equation}
u_k = u_{-k} \text{, } v_k = - v_{-k} \;, \text{  and  } \; u_k^2 + v_k^2 =1\;,
\end{equation}
where
\begin{equation}
\label{relation1}
u_k= \cos \Big(\frac{\phi_k}{2}\Big) \;, \text{    } 
v_k= \sin \Big(\frac{\phi_k}{2}\Big) \;,
\end{equation}
with $\phi_k$ given by
\begin{equation}
\label{relation2}
\tan (\phi_k) = \frac { 2 g \gamma \sin k} { -1 + 2g \cos k}\;.
\end{equation}

In this way, the quasiparticle creation and annihilation operators
$\gamma^\dagger_k$  and $\gamma^{\;}_k$, satisfy  the canonical
fermionic  anti-commutation relations of Eq. (\ref{anticom}), and the
Hamiltonian  may be finally rewritten as 
\begin{equation}
\label{diago}
H=  \sum\limits_{k \epsilon V} \xi_k (\gamma^{\dagger}_k \gamma^{\;}_k - 1/2)\;,
\end{equation}
where $\xi_k =2\sqrt {(-1+2g \cos k)^2 + 4 g^2 \gamma^2 \sin^2 k }$ is
the quasiparticle energy.  Since in general $\xi_k > 0$,  the ground
state is the quantum state with no quasiparticles (BCS state
\cite{Takahashi}),  such that $\gamma_k \ket{\sf BCS} =0$.  
Thus, one finds
\begin{equation}
\label{Ground}
\ket{\sf BCS} = \prod \limits_{k \epsilon V_+} (u_k + i v_k \tilde{c}^{\dagger}_k
\tilde{c}^{\dagger}_{-k} ) \ket{\sf vac}\;,
\end{equation}
where $\ket{\sf vac}$ is the state with no fermions ($\tilde{c}_k
\ket{\sf vac}=0$).  

Excited states with an even number of fermions ($K=+1$) can be
obtained applying pairs of quasiparticle creation operators
$\gamma^\dagger_k$ to the $\ket{{\sf BCS}}$ state. However, one should
be more rigorous when obtaining excited states with an odd number of
particles, since $K=-1$ implies periodic boundary conditions in
Eq. (\ref{Ham2}), and the new set of possible $k$'s (wave vectors) is
$\overline{V}=[ -\pi, \cdots, -\frac{2 \pi}{N}, 0, \frac{2 \pi}{N},
\cdots, \frac{2(N-1) \pi}{N} ]$ (different of $V$).

\subsection{QPT and critical point}

\begin{figure}[hbt]
\begin{center}
\includegraphics[width=11.5cm]{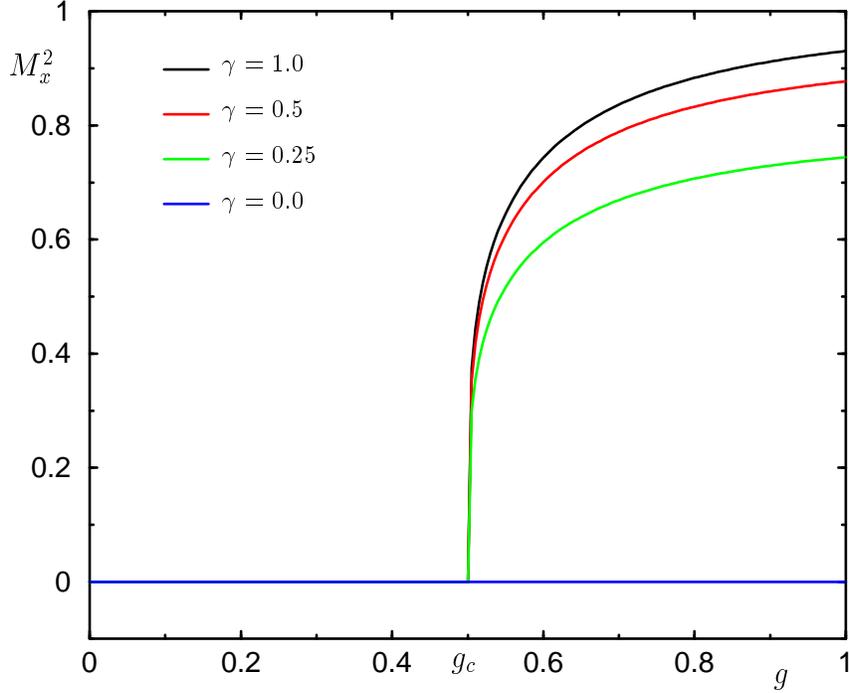}
\end{center}
\caption{Order parameter $M_x^2$ in the thermodynamic limit as a function of 
$g$ for different anisotropies $\gamma$.  The critical point is at $g_c=1/2$. }
\label{orderparameter}
\end{figure}

In Fig. \ref{orderparameter} we show the order parameter $M_x^2=
\lim_{N \rightarrow \infty} \langle \sigma_x^1 \sigma_x^{N/2} \rangle$
as a function of $g$ in the thermodynamic limit and for different
anisotropies $\gamma$ \cite{bamc}.  We observe that $M_x^2=0$ for $g 
\leq g_c$ and $M_x^2 \neq 0$ for $g>g_c$, so the critical point is
located at $g_c=1/2$, regardless of the value of $\gamma$.  The value
of $g_c$ can also be obtained by setting $\xi_k=0$ in Eq.
(\ref{diago}), where the gap vanishes. 

Notice that the Hamiltonian of Eq. (\ref{Ham1}) is invariant under the
transformation that maps $(\sigma_x^i;\sigma_y^j;\sigma_z^k) \mapsto (
-\sigma_x^i;-\sigma_y^j; \sigma_z^k)$ ($\zlxs_2$ symmetry), implying
that $\langle \sigma_x^i \rangle = 0$ for all $g$. However, since in
the thermodynamic limit the ground state becomes two-fold degenerate,
for $g>g_c$ , it is possible to build up a ground state where the discrete
$\zlxs_2$ symmetry is broken, i.e. $\langle \sigma_x^i \rangle \neq
0$.  This statement can be easily understood if we consider the case
of $\gamma=1$, where for $0\leq g < g_c$ the ground state has no
magnetization in the $x$ direction: For $g=0$, the spins align with
the magnetic field, while an infinitesimal spin interaction disorders
the system and $M_x=0$. On the other hand, for $g \rightarrow \infty$
the states $\ket{g_1}= \frac{1}{\sqrt{2}}[\ket{ \rightarrow , \cdots ,
\rightarrow} + \ket{ \leftarrow, \cdots, \leftarrow}]$ and $\ket{g_2}=
\frac{1}{\sqrt{2}}[\ket{ \rightarrow, \cdots , \rightarrow} - \ket{
\leftarrow, \cdots, \leftarrow}],$ with $\ket{\rightarrow}=
\frac{1}{\sqrt{2}} [\ket{\uparrow} +\ket{\downarrow}]$ and
$\ket{\leftarrow}= \frac{1}{\sqrt{2}} [\ket{\uparrow}
-\ket{\downarrow}]$ become degenerate in the thermodynamic limit, and
a ground state with $\langle \sigma_x^i \rangle \neq 0$ can be
constructed from a linear combination.

Remarkably, this paramagnetic-to-ferromagnetic QPT does not exist in the
isotropic limit ($\gamma=0$). In this case, the Hamiltonian of Eq. (\ref{Ham1})
has a continuous $\fu(1)$ symmetry; that is, it is invariant under any
$\hat{z}$ rotation of the form $\exp [i \theta \sum_j \sigma_z^j]$. Since the
model is one-dimensional, this symmetry cannot be spontaneously broken,
regardless of the magnitude of the coupling constants. Nevertheless, a simple
calculation of the ground state energy indicates a divergence in its second
derivative at the critical point $g_c=1/2$, thus, a second order non-broken
symmetry QPT. For $g < g_c$ all the spins (in the ground state) are aligned
with the external magnetic field, with total magnetization in the $\hat{z}$
direction $M_z=\sum_j \langle \sigma_z^j \rangle = -N$, and the quantum phase
is gapped. For $g \ge g_c$, the total magnetization in the $\hat{z}$ direction
is $M_z \ge -N$, the gap  vanishes, and the quantum phase becomes critical
(i.e., the spin-spin correlation functions decay with a power law), with an
emergent  $\fu(1)$ gauge symmetry \cite{advances}. Then, in terms of fermionic
operators (Eq. (\ref{Ham2})), an insulator-metal (or superfluid)  like second
order QPT exists at $g_c$ for the isotropic case, with no symmetry order
parameter. It is a Lifshitz transition.

\subsection{$\fu(N)$-purity in the BCS state, and critical behavior}

The $\ket{\sf BCS}$ state of Eq. (\ref{Ground}) is a GCS of the
algebra of observables $\fh=\fso (2N)$, spanned by an orthonormal 
Hermitian basis which is constructed by adjoining to the basis of 
$\fu(N)$ given in Eq. (\ref{uNbasis}) the following set $\fr$ of 
number-non-conserving fermionic operators:
\begin{equation}
\fr=\left\{
\begin{array}{ll}
(c^{\dagger}_j c^\dagger_{j'} + c^{\;}_{j'} c^{\;}_j) & \mbox{  with } 1\leq j<j' \leq N \cr
i(c^{\dagger}_j c^\dagger_{j'} - c^{\;}_{j'} c^{\;}_j) & \mbox{  with } 1\leq j<j' \leq N \cr
\end{array}
\right. \:, \hspace{5mm}  \fso(2N) = \fu(N) \oplus \fr \:.
\label{so2Nbasis}
\end{equation}
Then, the $\ket{\sf BCS}$ state is generalized unentangled with
respect  to the $\fso (2N)$ algebra and its purity $P_{\fh}$ (Eq.
(\ref{purity1})) contains no information about the phase transition:
$P_{\fh}=1\mbox{ } \forall g,\gamma$. Therefore, in order to
characterize the QPT we need to look at the possible subalgebras of
$\fso(2N)$.  A natural choice is to restrict to operators which
preserve the total fermion number that is, to consider the $\fu(N)$
algebra  defined in Section \ref{sectionunpurity}, relative to which
the $\ket{\sf BCS}$  state may become generalized entangled. (Note that
as mentioned in Section  \ref{sectionunpurity}, the  $\fu(N)$ algebra can
also be written in terms of the fermionic operators 
$\tilde{c}^\dagger_{k}$ and $\tilde{c}_{k}$, with $k$ belonging to the
set $V$.)

In the $\ket{\sf BCS}$ state, $\langle \tilde{c}^\dagger_{k}
\tilde{c}^{\;}_{k'}  \rangle \neq 0$ only if $k=k'$, thus using Eq.
(\ref{unpurity}) the purity  relative to $\fh=\fu(N)$ is:
\begin{equation}
\label{unpurity2}
P_{\fh} (\ket{\sf BCS}) = 
\frac{4}{N} \sum\limits_{k \epsilon V} \langle \tilde{c}^{\dagger}_k \tilde{c}^{\;}_k -1/2 \rangle^2
= \frac{4}{N} \sum\limits_{k \epsilon V} (v_k^2 -1/2)^2 \;,
\end{equation}
where the coefficients $v_k$ can be obtained from Eqs.
(\ref{relation1}) and (\ref{relation2}).  In particular, for $g=0$ the
spins are aligned with the magnetic field and the fully polarized
$\ket{\sf BCS}_{g=0}= \ket{\downarrow,  \downarrow, \ldots ,\downarrow}$
state is generalized unentangled in this limit (a GCS of $\fu(N)$ with
$P_{\fh}=1$). In the thermodynamic limit, the purity relative to the
$\fu(N)$ algebra can be obtained by integrating Eq. (\ref{unpurity2}):
\begin{equation}
P_{\fh}(\ket{\sf BCS})=\frac{2}{\pi}\int\limits_0^{2 \pi} (v_k^2 -1/2)^2 dk\:,
\end{equation}
leading to the following result:
\begin{equation}
\label{purity}
P_{\fh} (\ket{\sf BCS}) = \left\{ \begin{array}{cl} \frac {1}{1- \gamma ^2} 
\Big[ 1 - \frac{\gamma^2}{\sqrt{1 - 4g^2 (1-\gamma^2)}} \Big] 
& \text{ if } g \leq 1/2 \\  
\frac{1}{1+ \gamma} & \text{ if } g>1/2 \end{array} \right.\; .
\end{equation}
Although this function is continuous, its derivative is not and has a
drastic change at $g=1/2$, where the QPT occurs. Moreover, $P_{\fh}$ is
minimum for $g>1/2$ implying maximum entanglement at the transition
point and in the ordered  (ferromagnetic) phase.  Remarkably, for
$g>1/2$ and $N\rightarrow \infty$, where  the ground state of  the
anisotropic XY model in a transverse magnetic field is  two-fold
degenerate, $P_{\fh}$ remains invariant for arbitrary linear 
combinations of the two degenerate states.

\begin{figure}[ht]
\begin{center}
\includegraphics[width=14cm]{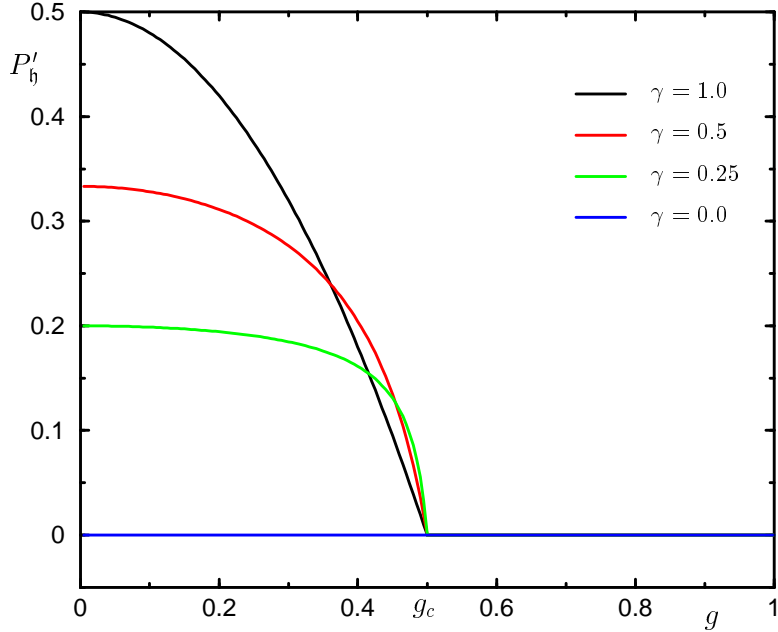}
\end{center}
\caption{Shifted purity $P'_{\fu(N)}$ of the $\ket{\sf BCS}$ as a
function of $g$  for different anisotropies $\gamma$, Eq.
(\ref{disorder0}). $P'_{\fu(N)}$ behaves like a disorder  parameter for
this model, sharply identifying the QPT at $g_c=1/2$. }
\label{figpur}
\end{figure}

As defined, for large $g$ the purity $P_{\fh}$ approaches a constant
value  which depends on $\gamma$.  It is convenient to remove such
dependence in the  ordered phase by introducing a new quantity
$P'_{\fh}= P_{\fh} - \frac{1}{1+\gamma}$ (shifted purity).  We thus
obtain
\begin{equation}
P'_{\fh}(\ket{\sf BCS}) = \left\{ \begin{array}{cl} \frac{\gamma}{1- \gamma^2} 
\Big[ 1 - \frac{\gamma} {\sqrt{1 - 4g^2 (1- \gamma^2)}} \Big]& 
\text{ if } g \leq 1/2 \\  0 & \text{ if } g>1/2  \end{array} \right. \;.
\label{disorder0}
\end{equation}
The new function $P'_{\fh}$ behaves like a {\it disorder parameter} for
the system,  being zero in the ferromagnetic (ordered) phase and
different from zero in the  paramagnetic (ordered) one.  The behavior
of $P'_{\fh}$ as a function of $g$ in  the thermodynamic limit is
depicted in Fig. \ref{figpur} for different values of  $\gamma$.  In
the special case of the Ising model in a transverse magnetic field 
($\gamma=1$), one has the simple behavior $P'_{\fh}= 1/2 -2g^2$  for $g
\leq 1/2$  and $P'_{\fh}=0$ if $g>1/2$.

\begin{figure}[hbt]
\begin{center}
\includegraphics[width=12cm]{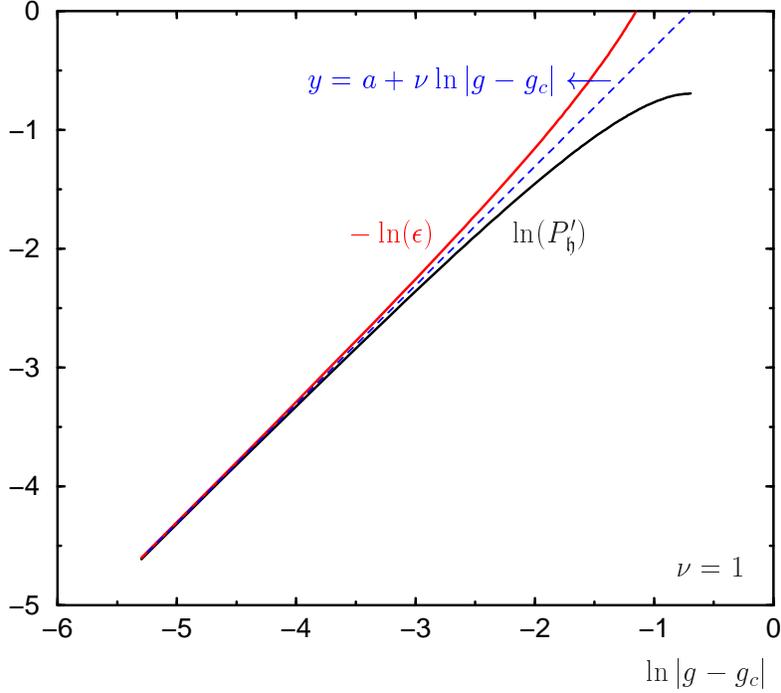}
\end{center}
\caption{Scaling properties of the disorder parameter for anisotropy
$\gamma=1$. The exponent $\nu=1$ belongs to the Ising universality class.}
\label{scaling}
\end{figure}

The critical behavior of the system is characterized by a power-law
divergence  of the {\it correlation length} $\epsilon$, which is
defined such that for $g<1/2$, $\lim_{|i-j| \rightarrow \infty} |
\langle \sigma_x^i \sigma_x^j \rangle | \sim \exp
(-\frac{|i-j|}{\epsilon})$. Thus, $\epsilon \rightarrow \infty$ signals
the emergence of long-range correlations in the ordered region $g>1/2$.
Near the critical point ($g \rightarrow 1/2^-$) the correlation length
behaves  as $\epsilon \sim (g_c - g)^{- \nu}$, where $\nu$ is a
critical exponent and the value $\nu =1$ corresponds to the Ising
universality class.  Let the  parameter $\lambda_2 = e^{-1/\epsilon}$. 
The fact that the purity contains  information about the critical
properties of the model follows from the  possibility of expressing
$P'_{\fh}$ for $g <1/2$ as a function of the  correlation length, 
\begin{equation}
\label{disorder}
P'_{\fh} (\ket{\sf BCS}) = 
\frac{\gamma} {1- \gamma^2} \left [ 1 + \frac {\gamma} {2g \lambda_2 (1 -
\gamma) -1 } \right ]
\end{equation}
where a known relation between $g$, $\gamma$, and $\lambda_2$ has been 
exploited \cite{bamc}.  Performing a Taylor expansion of Eq.
(\ref{disorder})  in the region $g \rightarrow 1/2^-$, we obtain $
P'_{\fh} \sim 2  (g_c - g )^\nu /\gamma $ with $\nu=1$ and $\gamma >0$ 
(Fig. \ref{scaling}). Thus, the name disorder  parameter for $P'_{\fh}$
is consistent.

\begin{figure}[hbt]
\begin{center}
\includegraphics[width=11cm]{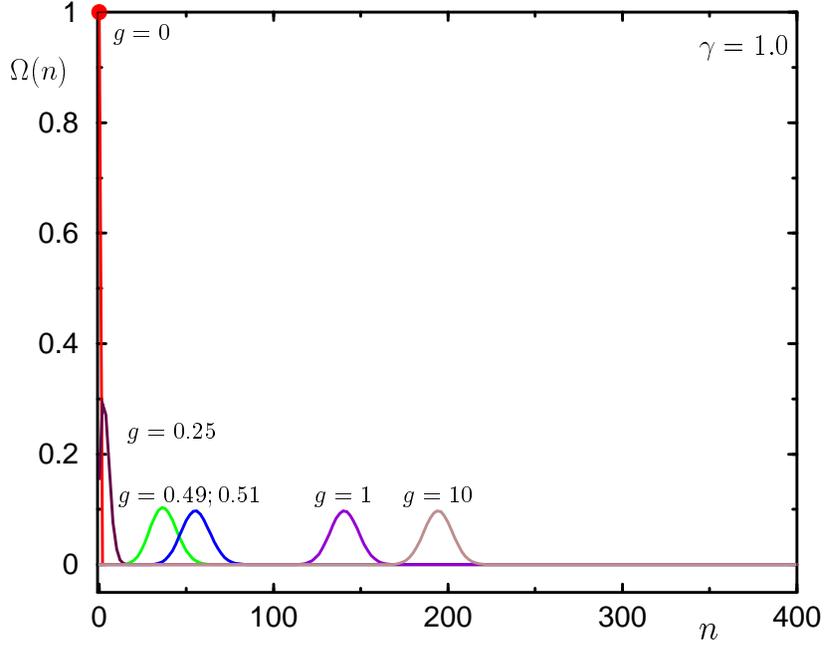}
\label{distribution}
\end{center}
\caption{Distribution of the fermion number in the $\ket{\sf BCS}$ state
for a chain of $N=400$ sites and anisotropy $\gamma=1$.}
\label{fluctuation}
\end{figure}

Some physical insight in the meaning of the ground-state purity may be
gained by  noting that Eq. (\ref{unpurity2}) can be written in terms
of the fluctuations  of the total fermion operator $\hat{N}$ 
\begin{equation}
\label{fluct1}
P_{\fh}(\ket{\sf BCS}) =1 - \frac{2}{N} \Big(\langle \hat{N}^2 \rangle - 
\langle \hat{N}\rangle ^2 \Big)\;.
\end{equation}
where the $\ket{\sf BCS}$-property  $\langle \tilde{c}^\dagger_k
\tilde{c}_{k'}\rangle= \delta_{k,k'} v_k^2$ has been used.   In
general, the purity relative to a given algebra can be written in terms
of fluctuations of observables \cite{baknorvi}. 
Since fluctuations of observables are at the root of QPTs it is not
surprising  that this quantity succeeds at identifying the critical
point. Interestingly,  by recalling that $P_{\fso(2N)}(\ket{\sf
BCS})=1$, the $\fu(N)$-purity can also be formally expressed as
\begin{equation}
\label{fluct2}
P_{\fu(N)}(\ket{\sf BCS}) =1 - \sum_{A_\alpha \in \fr} 
\langle A_\alpha \rangle ^2 \:,
\end{equation}
where the sum only extends to the non-number-conserving $\fso(2N)$-generators
belonging to the set $\fr$ specified in Eq. (\ref{so2Nbasis}). 
Thus, the purity is entirely contributed by expectations of operators
connecting different $\fu(N)$-irreps, the net effect of correlating
representations with a different particle number resulting in the
fluctuation of a {\it single} operator, given by $\hat{N}=\sum_k
\tilde{c}_k^\dagger \tilde{c}_k$.  In Fig. \ref{fluctuation}, we  show
the probability $\Omega(n)$ of having $n$ fermions in a chain of
$N=400$ sites  for $\gamma=1$.  We observe that for $g>1/2$ the
fluctuations remain almost constant,  and so does the purity.

Again, the isotropic case ($\gamma=0$) is particular in the sense that
$P_{\fh}=1$ (or $P'_{\fh}=0$, see Fig. \ref{figpur}), without
identifying the corresponding metal-insulator QPT. The reason is that
in this limit, the Hamiltonian of Eq. (\ref{Ham2}) contains only
fermionic operators that preserve the number of particles (i.e., $H
\in \fu(N)$), and the ground state of the system is always a GCS of
the $\fu(N)$ algebra. Therefore, in order to obtain information about
this QPT, one should look into algebras other than $\fu(N)$, relative 
to which the ground state is generalized entangled. For example, in
Sec. \ref{localalgebra} we study the purity relative to the local
algebra of observables and in Fig. \ref{figpur2} we show that it
succesfully identifies the QPT in the isotropic case, being maximum
for $g \le g_c$ (thus implying generalized unentanglement).

\subsection{ Comparison with concurrence}

\begin{figure}[t]
\begin{center}
\includegraphics[width=10cm]{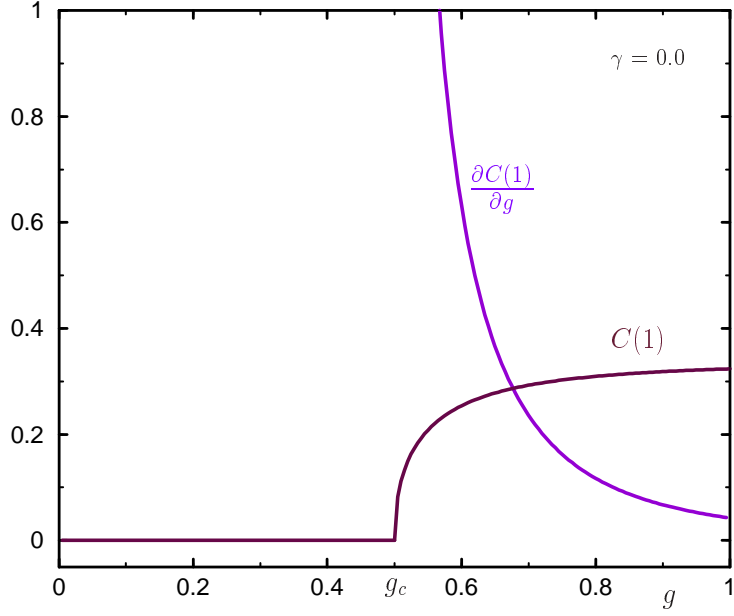}
\end{center}
\caption{Nearest-neighbor concurrence and its derivative for the $\ket{\sf BCS}$ 
state as a function of $g$ in the isotropic XY model, $\gamma= 0$. Both curves
correspond to the exact solution in the thermodynamic limit.  The value of 
$\partial  C(1)/ \partial g$ below $g_c$ is also zero as $C(1)$ (not shown). 
}
\label{isoconc}
\end{figure}

As mentioned, the critical behavior of the XY model in a transverse
field  has also been investigated by looking at various quantities
related to the  concurrence, which is intrinsically a measure of
bipartite entanglement.  For a generic mixed state $\rho$ of two
qubits, the latter is calculated  as \cite{wo}
$$ C (\rho) = \text{max} \{ \lambda_1-\lambda_2- \lambda_3-\lambda_4,
0 \} \:, $$ where $\lambda_1 \geq \ldots\geq \lambda_4$ are the square
roots of the eigenvalues of the matrix $R=\rho \tilde{\rho}$ and
$\tilde{\rho}= \sigma_y \otimes \sigma_y \rho^\ast \sigma_y \otimes
\sigma_y$.  The concurrence for the reduced density operator
$\rho_{\ell,m}$ of two nearest-neighbor qubits ($|\ell -m|=1$) and
next-nearest-neighbor ($|\ell -m|=2$) qubits on a lattice has been
investigated in detail in Ref. \cite{osamfafa}.  Since, thanks to
translational invariance, $\rho_{\ell,m}$ depends on the qubit indexes
only via their distance, we will use the notation $C(1)$, $C(2)$ for
the resulting quantities as in \cite{osamfafa}.  While the results
reported in the above work nicely agree with the scaling behavior
expected for this model, the emerging picture based on concurrence
cannot be regarded as fully satisfactory.  As
stressed in \cite{osamfafa}, the entanglement as quantified by
the nearest-neighbor concurrence is {\em not} directly an indicator of
the QPT in this model, 
showing maximum entanglement at a point which is not related
to the QPT.  
However, the derivative $\partial C(1)/\partial
g$ of the concurrence with respect to the spin-spin coupling parameter
can be seen to diverge logarithmically at the critical point for
$\gamma >0$, and with a power law for the isotropic case 
[Fig. (\ref{isoconc})], identifying the critical point in this model. 
Such a divergence is not found when
analyzing, at the isotropic point, other QPTs in models of interest, 
like the one-dimensional 
anisotropic Heisenberg chain (see, for instance, \cite{glas}).
Therefore, it suggests that the identification of
a critical point using concurrence could be a hard task in general.

\subsection{Purity of the BCS state relative to the local algebra }
\label{localalgebra}

\begin{figure}[ht]
\begin{center}
\includegraphics[width=13.5cm]{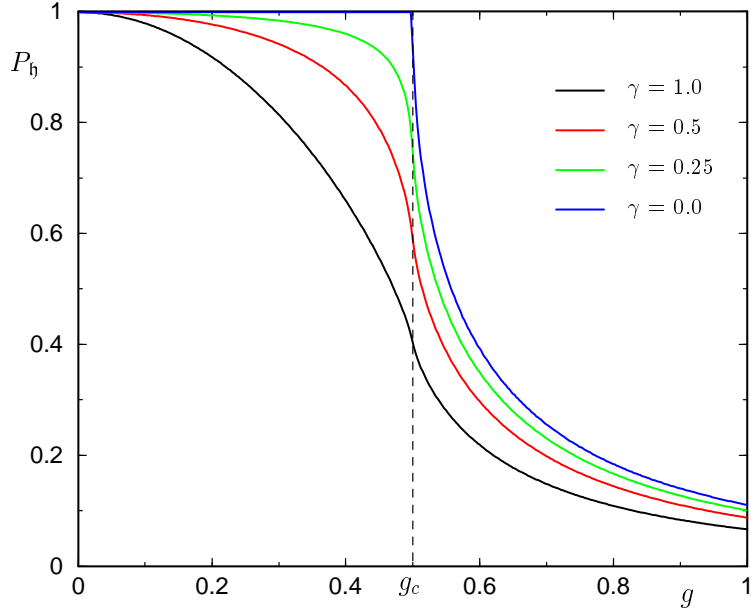}
\end{center}
\caption{Purity of the $\ket{\sf BCS}$ state relative to the local algebra
$\bigoplus \limits_{i=1}^N \fsu(2)_i$, as a function of $g$ for different
anisotropies $\gamma$ ($g_c=1/2$). The number of sites $N=400$ as in 
Fig. \ref{fluctuation}. }
\label{figpur2}
\end{figure}

Finally, we have also investigated the behavior of the purity of the
$\ket{\sf BCS}$ state relative to the algebra of local observables
$\fh=\bigoplus\limits_{i=1}^N \fsu(2)_i$.  Using Eq. (\ref{purity3}),
this is physically related to the total magnetization $M^2_z$ along
$z$.  The resulting behavior is plotted in Fig. \ref{figpur2} as a
function of $g$ and $\gamma$.  As explained in Section
\ref{Nspinexamples}, this is a measure of the usual notion of
entanglement in the $N$-spin-1/2s system.  In particular, the
$\ket{\sf BCS}$ state is unentangled for $g\rightarrow 0$ (where
$\ket{\sf BCS} \sim \ket{-\frac{1}{2}}_1 \otimes \cdots \otimes
\ket{-\frac{1}{2}}_N$), thus $P_{\fh} \rightarrow 1$ in this limit.
Moreover, for $g \rightarrow \infty$ we have $\ket{\sf BCS} \sim
\ket{{\sf GHZ}_{\frac{1}{2}}^N}$ (up to local rotations), thus
$\ket{\sf BCS}$ becomes maximally entangled, and $P_{\fh} \rightarrow
0$.

\begin{figure}[tb]
\begin{center}
\includegraphics[width=11.5cm]{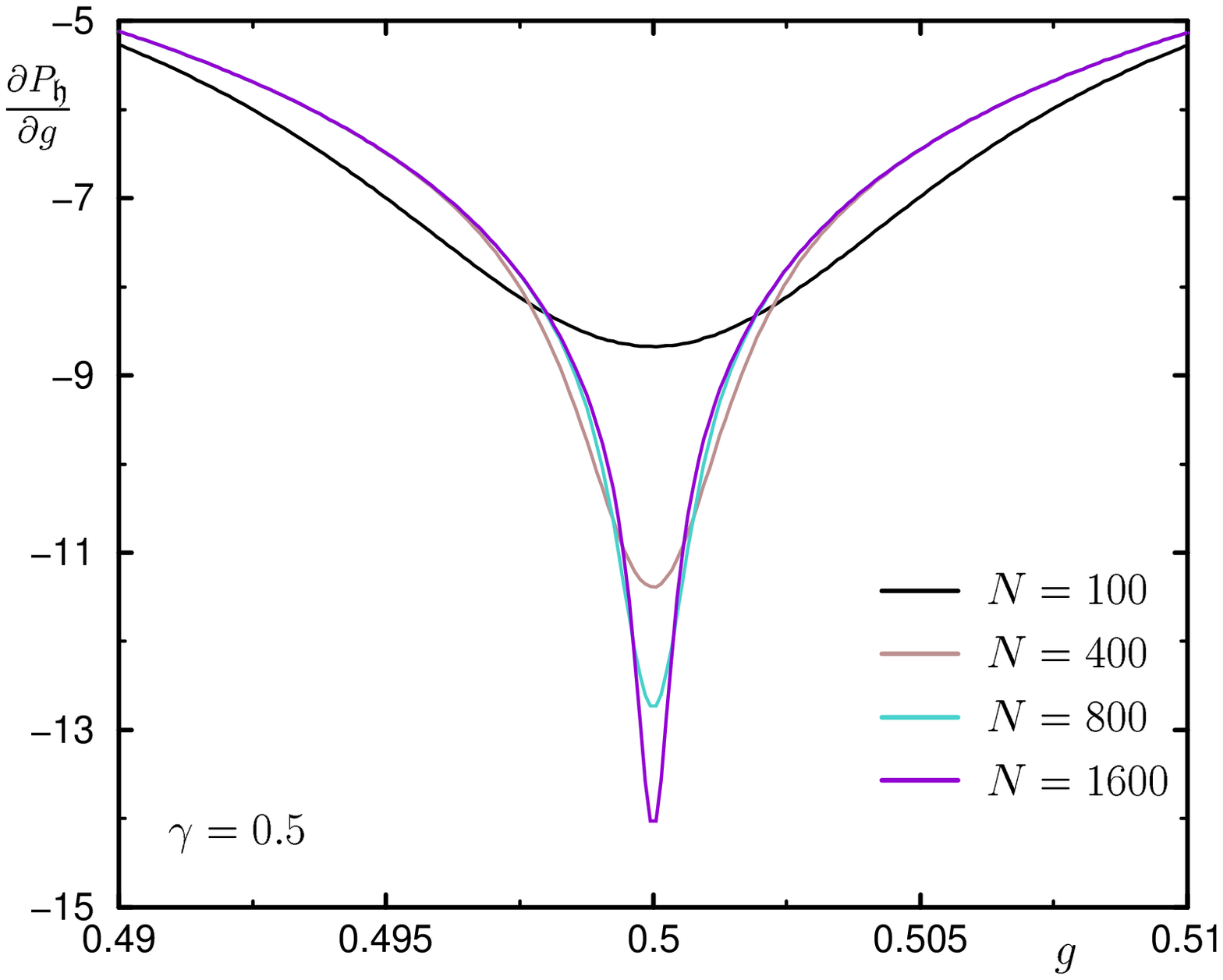}
\end{center}
\caption{Derivative of the purity of the $\ket{\sf BCS}$ state relative to the
local algebra as a function of $g$ for $\gamma= 0.5$ and different lattice sizes.}
\label{derivative}
\end{figure}

Compared to the purity relative to the $\fu (N)$ algebra, the purity relative
to $\fh=\bigoplus\limits_{i=1}^N \fsu(2)_i$ is not as good an indicator of the
phase transition when $\gamma>0$,  in the sense that it does not present 
any drastic change in its behavior. However, its derivative with respect
to the spin-spin coupling parameter diverges at the critical point in this
model [Fig. \ref{derivative}]. Only in the
isotropic case ($\gamma=0$) the purity relative to the local algebra presents a
drastic change at the critical point (see Fig. \ref{figpur2}). In this case,
the operator $M_z=\frac{1}{N}\langle \sum_j \sigma_z^j \rangle$ 
for $g\rightarrow g_c^+$ scales as
\begin{equation}
 M_z+1 \sim  (g-g_c)^\chi 
\end{equation}
with the exponent being $\chi=1/2$. On the other hand,
this exponent can also be obtained from
the purity relative to the local algebra, in the same limit:
\begin{equation}
1- P_{\fh} \sim  (g-g_c)^\chi.
\end{equation}
Therefore, this measure of entanglement is also a good indicator of the QPT
for the isotropic case.


\section{Conclusions}

In this paper, we have explored the usefulness of generalized
entanglement (GE) for characterizing the broken (and one example of
non-broken) symmetry quantum phase transitions (QPTs) present in
different lattice systems. As we focused on situations where the
physically relevant observables form a Lie algebra, a natural GE
measure provided by the relative purity of a state relative to the
algebra has been used to identify and characterize these transitions.

In Sections \ref{spinexamples} and \ref{Nspinexamples}, we showed using
several illustrative examples how the concept of $\fh$-purity can be
useful for different spin systems, by encompassing the usual notion of
entanglement if the family of all local observables is distinguished.  
In addition, the possibility to directly apply the GE notion to arbitrary
quantum systems, including indistinguishable particles, was explicitly
shown in Section \ref{sectionunpurity}, using fermionic systems as a
relevant case study. Depending on the subset of observables chosen, the
$\fh$-purity contains information about different $n$-body correlations
present in the quantum state, allowing for a more general and complete
characterization of entanglement. Finally, in Sections \ref{lmg} and
\ref{xymodel} we showed that the $\fh$-purity successfully
distinguishes between the different phases present in two lattice
systems, where the critical points are characterized by a broken
symmetry (or non-broken symmetry in the case of the isotropic XY model 
in an external magnetic field) and the usual notion of entanglement cannot be
straightforwardly applied. As also discussed in Section
\ref{purity-QPT}, the most critical step is to determine which subset
of observables may be relevant in each case, since the $\fh$-purity
must contain information about the quantum correlations that play a
dominant role in the QPT. In particular, the ground state of the two
models we considered can be exactly calculated and the relevant quantum
correlations in the different phases are well understood, thus choosing
this subset of observables becomes relatively easy.  

Applying these concepts to a more general case, where the ground state
of the system cannot be exactly computed, can be done in principle by
following the same strategy. However, determining in a systematic way
the minimal subset of observables $\fh$ whose purity is able to signal 
and characterize the QPT, thereby providing the relevant correlations,
requires an elaborate analysis.  Even more interesting, perhaps, is the
open question of finding the minimal number of GE measures, possibly
including measures of GE relative to different observable sets, needed
to unambiguously characterize the universality class of a transition,
obtaining all of its critical exponents.  Finally, a fascinating 
direction for further investigation is to explore the significance of 
the GE notion within topological quantum-information settings~\cite{kitaev}
and to understand what generalizations might be needed to handle 
topological QPTs.

\acknowledgments
It is a pleasure to thank James Gubernatis, Leonid Gurvits, Juan Pablo
Paz, and Wojciech Zurek for discussions.  We acknowledge support from
the US DOE through Contract No. W-7405-ENG-36. H.~B. and L.~V. gratefully
acknowledge financial support from the Los Alamos Office of the Director.

\appendix

\section{Separability, Generalized Unentanglement, and Local Purities}
\label{app1}

Given a quantum system ${\cal S}$ whose states $\ket{\psi}$ belong to 
a Hilbert space ${\cal H}$ of dimension dim$({\cal H})=d$, the purity 
relative to the (real) Lie algebra of all traceless observables $\fh=\fsu(d)$
spanned by an orthogonal, commonly normalized Hermitian basis 
$\{ A_1 \cdots A_{L} \}$, $L=d^2-1$, is, according to Eq. (\ref{purity1}), 
given by:
\begin{equation}
\label{puritysubsystem}
P_{\fh} (|\psi\rangle ) = {\sf K} \sum\limits_{\alpha=1}^{L} \langle A_{\alpha}
\rangle^2 .
\end{equation}
The normalization factor ${\sf K}$ depends on $d$ and is determined so that 
the maximum purity value is 1. If {\sf Tr}$(A_\alpha A_\beta)=\delta_{\alpha, 
\beta}$ (as for the standard spin-$1$ Gell-Mann matrices), then
${\sf K}=d/(d-1)$, whereas in the case {\sf Tr}$(A_\alpha A_\beta)= d
\delta_{\alpha, \beta}$ (as for ordinary spin-$1/2$ Pauli matrices), 
${\sf K}=1/(d-1)$.  Recall that any quantum state $\ket{\psi} \in {\cal H}$ 
can be obtained by applying a group operator $U$ to a reference state 
$\ket{{\sf ref}}$ (a highest or lowest weight state of $\fsu (d)$); that is
\begin{equation}
\ket{\psi} =U \ket{{\sf ref}} \:,
\end{equation}
with $U=e^{i \sum_{\alpha} t_{\alpha} A_{\alpha}}$,  
and $t_{\alpha}$ real numbers.  Therefore, any quantum state $\ket{\psi}$ is 
a GCS of $\fsu (d)$, thus generalized unentangled relative to the 
algebra of all observables: $P_{\fh} (\ket{\psi}) =1$ for all $\ket{\psi}$.

Let now assume that ${\cal S}$ is composed of $N$ distinguishable susbsytems, 
corresponding to a factorization ${\cal H} = \bigotimes_{j=1}^N {\cal H}_j$, 
with dim$({\cal H}_j)=d_j$, $d=\prod_j d_j$. Then the set of all {\it local} 
observables on ${\cal S}$ becomes 
$\fh=\fh_{loc} = \bigoplus_j \fsu(d_j)$.  An orthonormal basis which is suitable
for calculating the local purity $P_{\fh}$ may be obtained by considering a 
collection of orthonormal bases 
$ \{  A^j_{\alpha_1} \cdots A^j_{\alpha_{L_j}} \}$, 
$L_j=d_j^2-1$, each acting on the $j$th subsystem that is, 
\begin{equation}
\label{Basis}
A^j_{\alpha_j} = \overbrace{ \one^1 \otimes \one^2 \otimes \cdots \otimes
\underbrace{A_{\alpha_j}}_{j^{th}\ \mbox{factor}} \otimes \cdots
\otimes \one^N}^{N\ \mbox{factors}} \: ,
\end{equation}
where $\one^j =\one/\sqrt{d_j}$. Then for any pure state $|\psi\rangle 
\in {\cal H}$ one may write
\begin{equation}
\label{puritysub}
P_{\fh} (|\psi\rangle )= {\sf K'} \sum\limits_{j=1}^N \Big[
\sum\limits_{\alpha_j=1}^{L_j} 
\langle A^j_{\alpha_j} 
\rangle^2 \Big] \:.
\end{equation}
By letting $\fh_j=\text{span}\{ A_{\alpha_j} \}$ be the Lie algebra of traceless
Hermitian operators acting on ${\cal H}_j$ alone, the above equation also 
is naturally rewritten as
\begin{equation}
\label{puritysub1}
P_{\fh} (|\psi\rangle )= {\sf K'} \sum\limits_{j=1}^N 
{1 \over {\sf K}_j} P_{\fh_j} (|\psi\rangle )\:, 
\hspace{5mm} {\sf K}_j ={d_j \over {d_j-1}}\:.
\end{equation}
The $\fh_j$-purity $P_{\fh_j}$ may be simply related to the conventional 
subsystem purity.  Let $\rho_j ={\sf Tr}_{i \not = j} 
(\{ |\psi\rangle\langle \psi|\})$ be the reduced density operator 
describing the state of the $j$th subsystem. Because the 
latter can be represented as 
\begin{equation}
\rho_j = \frac{\one}{d_j} + \sum\limits_{\alpha_j=1}^{L_j} 
\langle A_{\alpha_j} \rangle A_{\alpha_j} = 
\sum\limits_{\alpha_j=1}^{L_j} 
\langle A^j_{\alpha_j} \rangle A_{\alpha_j} \:,
\end{equation}
one can also equivalently express Eq. (\ref{puritysub}) as
\begin{equation}
\label{puritysub2}
P_{\fh} (|\psi\rangle )= {\sf K'} \sum\limits_{j=1}^N \Big[
{\sf Tr} \rho_j^2 - {1 \over d_j}  \Big] \:,
\end{equation}
that is, $P_{\fh_j} (|\psi\rangle ) = (d_j {\sf Tr} \rho_j^2 - 1)/(d_j-1) $. 
Clearly, the maximum value of either Eqs. 
(\ref{puritysub1}) or (\ref{puritysub2}) 
will be attained when, and only when, each of the conventional purities 
${\sf Tr}\rho_j^2=1$ $\leftrightarrow$ $P_{\fh_j} =1$ for all $j$, 
which allows determining the ${\sf K}'$-normalization factor as
\begin{equation}
{\sf K}'= \frac{1}{\sum_j {1 \over {\sf K}_j} } =
\frac{1}{ N - \sum_j {1 \over d_j}  } = 
\frac{1}{ N \Big( 1 - {1\over N} \sum_j {1 \over d_j} \Big) }  \:.
\end{equation}
Accordingly, 
\begin{equation}
P_{\fh_{loc}} (\ket{\psi}) = \text{max} =1 \leftrightarrow \ket{\psi} = 
\ket{\phi_1} \otimes \cdots \otimes \ket{\phi_N} \:,
\end{equation}
and the equivalence with the standard notions of separability and 
entanglement are recovered. 
Note that for the case of $N$ qubits considered in Section IIIB, the above 
value simplifies to ${\sf K}'=2/N$ which in turn gives the purity expression 
of Eq. (\ref{purity3}) once the standard unnormalized Pauli matrices are 
used ($A^j_{\alpha_j} = \sigma^j_{\alpha_j} /\sqrt{2}$, thus removing the 
overall factor 2).

\section{Cluster and Valence Bond Solid states are maximally entangled}
\label{appb}

In Ref. \cite{brra}, Briegel and Raussendorf introduced the so-called
cluster states for a system of $N$ qubits in $D$ space dimensions
which, in  the computational basis, are expressed as
\begin{equation}
\label{cluster}
\ket{\Phi}_C = \frac{1}{2^{N/2}} \bigotimes\limits_{j \in C} \left(
\ket{\uparrow}_j \prod\limits_{\gamma \in \Gamma}
\sigma_z^{(j+\gamma)} +  \ket{\downarrow}_j \right ),
\end{equation}
where $C$ defines the cluster ($C \subset \mathbb{Z}^D$) and $\gamma$
denotes some nearest neighbor qubits in the cluster: $\Gamma=\{ 1\}$
for $D=1$, $\Gamma= \{ (1,0),(0,1) \}$ for $D=2$, $\Gamma= \{
(1,0,0),(0,1,0) ,(0,0,1)\}$ for $D=3$, etc. We consider
$\sigma_z^{(j+\gamma)} \equiv 1$ when $j+ \gamma$ is not in $C$.

The usual notion of entanglement, as applied to a cluster state, is
recovered when the $\fh$-purity is calculated relative to the local
algebra $\fh = \bigoplus\limits_{j \in C}  \fsu (2)_j$ (see Appendix
\ref{app1}). For this purpose, we first calculate the expectation
values $\langle \sigma_\alpha^j \rangle_C$, with $\alpha=x,y,z$. One
can immediately realize that  $\langle \sigma_y^j \rangle_C =0$,
$\forall j$, since $\sigma_y^j$ is an  Hermitian operator (i.e.,
$\langle \sigma_y^j \rangle \in \mathbb{R}$) that acting on the $j$-th
qubit's state (in the natural basis) introduces a phase factor $\pm i$,
and the coefficients of Eq. (\ref{cluster}) are all real. Moreover, 
$\langle \sigma_z^j \rangle_C =0$, $\forall j$, since the weight
of every state of the natural basis is the same in Eq. (\ref{cluster}).
In other words, we have a linear combination of basis states where each
single qubit has the same probability of pointing up or down. Finally,
one can also prove that $\langle \sigma_x^j \rangle_C =0$, $\forall j$.
This can be done by using the eigenvalue equations $K_j \ket{\Phi}_C =
\pm \ket{\Phi}_C$, for the family of operators $K_j = \sigma_x^j
\prod\limits_{\gamma \in \bar{\Gamma}} \sigma_z^{(j + \gamma)}$, where
$\bar{\Gamma}=\Gamma \bigcup -\Gamma$ denotes  the set of all
nearest-neighbor  qubits to the $j$-th qubit. Therefore,  $\langle
\sigma_x^j \rangle_C = \pm \langle \sigma_x^j  K_j \rangle_C =\pm
\langle \prod\limits_{\gamma \in \bar{\Gamma}} \sigma_z^{(j + \gamma)}
\rangle_C$. Again, since Eq. (\ref{cluster}) is a  combination of all
the states of the computational basis with the same probability, we obtain
$\langle \sigma_x^j \rangle_C =0$. In this way, the $\fh$-purity (Eq.
(\ref{purity3})) is $P_{\fh}=0$, and the cluster states are maximally 
entangled relative to the local set $\fh = \bigoplus\limits_{j \in C} 
\fsu (2)_j$.

Another important class of spin states is the one defined by the
so-called {\em Valence Bond Solid} (VBS) states. These states have
been introduced in the context of Heisenberg-like magnets, and have 
been recently revisited in the context of quantum computation 
\cite{vbs}.  Their general form is 
\begin{equation}
\label{VBS}
\ket{\Phi}_{\sf VBS} =\prod_{\langle i,j \rangle} \left(
a^\dagger_i b^\dagger_j - b^\dagger_i a^\dagger_j \right )^M \ket{0},
\end{equation}
where $\langle i,j \rangle$ represent nearest-neighbor bonds of a
$D$-dimensional lattice of coordination ${\sf z}$, $a^\dagger_j$ and
$b^\dagger_j$ are Schwinger-Wigner boson (creation) operators on site
$j$ whose relation to $\fsu(2)$ spin-$S$ generators is
\begin{equation}
\label{Schwinger}
S_x^j=\frac{1}{2} (a^\dagger_j b^{\;}_j + b^\dagger_j a^{\;}_j), \ \ 
S_y^j=\frac{1}{2i} (a^\dagger_j b^{\;}_j - b^\dagger_j a^{\;}_j), \ \
S_z^j=\frac{1}{2} (a^\dagger_j a^{\;}_j - b^\dagger_j b^{\;}_j),
\end{equation}
with the constraint $a^\dagger_j a^{\;}_j + b^\dagger_j b^{\;}_j=2 S$, 
and $M=2S/{\sf z}$. $M$ being an integer makes the possible values
of $S$ to depend upon the connectivity of the lattice, which is 
defined by ${\sf z}$.  

We start by showing that the bond operators $a^\dagger_i b^\dagger_j -
b^\dagger_i a^\dagger_j$ are invariant under global spin rotations. 
The Schwinger-Wigner boson operators transform as vectors for $\fsu(2)$
rotations
\begin{equation}
\label{swtransform}
\pmatrix {a^\dagger_j \cr b^\dagger_j \cr} \rightarrow  
U_j \pmatrix {a^\dagger_j \cr b^\dagger_j \cr} U_j^\dagger =
\pmatrix {\cos \frac{\theta_2}{2} \ e^{i (\theta_3+\theta_1)/2} &
\sin \frac{\theta_2}{2} \ e^{i (\theta_3-\theta_1)/2} \cr 
-\sin \frac{\theta_2}{2} \ e^{-i (\theta_3-\theta_1)/2} & 
\cos \frac{\theta_2}{2} \ e^{-i (\theta_3+\theta_1)/2} \cr } 
\pmatrix {a^\dagger_j \cr b^\dagger_j \cr} 
\end{equation}
under an arbitrary spin rotation on lattice site $j$, defined by 
\begin{equation}
U_j = e^{i \theta_1 S_z^j} \ e^{i \theta_2 S_y^j} \ e^{i \theta_3
S_z^j} \:, \hspace{5mm}U_j U_j^\dagger=U_j^\dagger U_j = \one \:.
\end{equation}
Then, we can use this result to prove that
\begin{equation}
U_j U_i (a^\dagger_i b^\dagger_j - b^\dagger_i a^\dagger_j) U_i^\dagger
U_j^\dagger = a^\dagger_i b^\dagger_j - b^\dagger_i a^\dagger_j \:,
\end{equation}
implying, for $U^\dagger=\prod_j U_j^\dagger$,
\begin{equation}
U^\dagger \ \ket{\Phi}_{\sf VBS}=\ket{\Phi}_{\sf VBS}\: .
\end{equation}
Therefore, $\ket{\Phi}_{\sf VBS}$ belongs to the singlet irrep of the 
total spin $J_\alpha=\sum_j S_\alpha^j$ (i.e., $\langle J_\alpha 
\rangle_{\sf VBS} =0)$.

We want to show now that $\langle S_\alpha^j\rangle_{\sf VBS}=0,
\forall j$. We first observe that $\langle S_z^j\rangle_{\sf VBS}=0,
\forall j$, because the transformation that maps $a^\dagger_j
\mapsto b^\dagger_j$ and $b^\dagger_j \mapsto -a^\dagger_j$
(i.e., a global spin rotation about the $y$-axis, setting
$\theta_1=\theta_3=0$ and $\theta_2= \pi$ in Eq. (\ref{swtransform}))
implies $\langle a^\dagger_j a^{\;}_j \rangle_{\sf VBS} = \langle
b^\dagger_j b^{\;}_j \rangle_{\sf VBS}$.  Then, from the invariance
under global spin rotations and the singlet nature of $\ket{\Phi}_{\sf
VBS}$, we obtain $\langle S_x^j\rangle_{\sf VBS}=0=\langle
S_y^j\rangle_{\sf VBS}, \forall j$. In other words, the purity relative
to the algebra $\fh = \bigoplus\limits_{j}  \fsu (2)_j$ vanishes, 
meaning that  $\ket{\Phi}_{\sf VBS}$ is maximally
generalized entangled relative to this algebra. 

However, in order to make contact with the standard notion of
entanglement (Appendix \ref{app1}) we need to address the GE
relative to the algebra $\fh = \bigoplus\limits_{j}\fsu
(2S+1)_j$, that is, relative to the set of all local observables. For
simplicity, we only discuss the 1$D$ case for $S=1$ (i.e., $M=1$ in
Eq. (\ref{VBS})) but the reader could use the same techniques to
obtain results in higher $D$ dimensions and spin magnitude $S$.

The algebra $\fh = \bigoplus\limits_{j}\fsu (3)_j=\{ {\cal S}^j_{\mu\nu}
\}$,  
\begin{equation}
[{\cal S}^{j}_{\mu \mu'},{\cal S}^{j'}_{\nu \nu'}]=\delta_{j j'} \  
(\delta_{\mu'\nu} \ {\cal S}^{j}_{\mu \nu'}-\delta_{\mu \nu'} \ 
{\cal S}^{j}_{\nu \mu'} ) , 
\label{conm3}
\end{equation}
can be written in terms of the $\fsu(2)$ generators as \cite{advances}
\begin{eqnarray}
{\cal S}^j_{0 0}&=&\frac{2}{3}-(S_z^j)^2,\;\;{\cal S}^j_{1
1}=\frac{S_z^j(S_z^j+1)}{2}-\frac{1}{3} \ , \nonumber \\
{\cal S}^j_{1 0}&=&\frac{1}{2\sqrt{2}} [ S_+^j + \left \{ S_+^j,S_z^j \right \}
] \, \nonumber \\
{\cal S}^j_{0 1}&=&\frac{1}{2\sqrt{2}} [ S_-^j + \left \{ S_-^j,S_z^j \right \}
] \, \nonumber \\
{\cal S}^j_{2 0}&=&\frac{1}{2\sqrt{2}} [ S_-^j - \left \{ S_-^j,S_z^j \right \}
] \, \nonumber \\
{\cal S}^j_{0 2}&=&\frac{1}{2\sqrt{2}} [ S_+^j - \left \{ S_+^j,S_z^j \right \}
] \, \nonumber \\
{\cal S}^j_{1 2}&=& \frac{i}{2} \left \{ S_x^j,S_y^j \right \}+(S_x^j)^2 +
\frac{1}{2}(S_z^j)^2-1 \ ,  \nonumber \\
{\cal S}^j_{2 1}&=& \frac{1}{2i} \left \{ S_x^j,S_y^j \right \}+(S_x^j)^2+
\frac{1}{2}(S_z^j)^2-1 \ ,
\label{igene}
\end{eqnarray}
with $S^j_{\pm} = S^j_x \pm i S^j_y$. From the spin-rotational
invariance of the state $\ket{\Phi}_{\sf VBS}$ we get $\langle
(S_x^j)^2 \rangle_{\sf VBS}=  \langle (S_y^j)^2 \rangle_{\sf VBS}= 
\langle (S_z^j)^2 \rangle_{\sf VBS}= \frac{S(S+1)}{3}$ and, since
$S=1$, we obtain $\langle {\cal S}^j_{0 0} \rangle_{\sf VBS} = \langle 
{\cal S}^j_{1 1} \rangle_{\sf VBS} =0$. Moreover, the spin-rotational
invariance also implies that $\langle S_\alpha^j S_{\alpha '}^j
\rangle_{\sf VBS}$ remains the same constant $\forall \alpha \neq
\alpha'$. Then, for example, applying a global $\pi$-rotation about
the $y$-axis to $\ket{\Phi}_{\sf VBS}$ (i.e., the operation that maps
$S_z^j \mapsto -S_z^j$ and $S_y^j \mapsto S_y^j$) we obtain
$\langle S_y^j S_z^j \rangle_{\sf VBS} = -\langle S_y^j S_z^j
\rangle_{\sf VBS}=0$, hence, $\langle {\cal S}^j_{\mu\nu} \rangle_{\sf
VBS} =0$. Therefore, the state $\ket{\Phi}_{\sf VBS}$ ($S=1, M=1$) is
maximally entangled when using the standard notion of entanglement
($P_{\fh}=0$, for the algebra of all local observables $\fh =
\bigoplus\limits_{j}\fsu (3)_j$).

\section{Relation between Purity in the 
local algebra and the Meyer-Wallach measure of entanglement}
\label{app2}

In Ref. \cite{mewa}, Meyer and Wallach define a measure of entanglement $Q$ 
for pure states of qubit systems, that is invariant under local unitary operations 
(local rotations). For this purpose, they first define the mapping $l_j(b)$ acting 
on product states as
\begin{equation}
l_j(b) \ket{b_1, \cdots ,b_N} = \delta_{bb_j} \ket {b_1 ,\cdots ,\hat{b}_j,
\cdots ,b_N},
\end{equation}
where $b$ and $b_j$ are either the states $\ket{\frac{1}{2}}$ or
$\ket{-\frac{1}{2}}$, and $\hat{b}_j$ denotes the absence of the $j$-th qubit.
On the other hand, any $N$-qubits pure  quantum state can be written in the
natural basis ($z$-component of  the spin equal to $\pm \frac{1}{2}$) as
\begin{equation}
\ket{\psi}=\sum\limits_{i=1}^{2^{N-1}} \Big[g_i^j \ket{\frac{1}{2}}_j + h_i^j
\ket{-\frac{1}{2}}_j \Big] \ket{\phi_i},
\end{equation}
where $g_i^j$ and $h_i^j$ are complex coefficients, and the orthonormal states
$\ket{\phi_i}$ of $N-1$ qubits (absence of the $j$-th qubit) are also written
in the natural basis. Therefore, the action of $l_j(b)$ on $\ket{\psi}$ is
\[
\begin{array}{l}
l_j(\frac{1}{2}) \ket{\psi}= \sum\limits_{i=1}^{2^{N-1}} g_i^j \ket{\phi_i} \\
l_j(-\frac{1}{2}) \ket{\psi}= \sum\limits_{i=1}^{2^{N-1}} h_i^j \ket{\phi_i}. 
\end{array}
\]

Then, they define the entanglement $Q (\ket{\psi})$ as
\begin{equation}
\label{meyer}
Q(\ket{\psi}) = \frac{4}{N} \sum\limits_{j=1}^N D\Big(l_j(\frac{1}{2}) 
|\psi \rangle , l_j(-\frac{1}{2}) |\psi \rangle \Big)\;,
\end{equation}
where the distance between two quantum states $\ket{u}=\sum u_i \ket{\phi_i}$
and $\ket{v}= \sum v_i \ket{\phi_i}$ is
\begin{equation}
D(u,v) = \frac{1}{2} \sum\limits_{i,j} |u_iv_j - u_jv_i|^2\;.
\end{equation}
Therefore,
\begin{equation}
D\Big(l_j(\frac{1}{2}) |\psi \rangle , l_j(-\frac{1}{2}) |\psi\rangle \Big) = 
\frac{1}{2} \sum\limits_{i,i'}
| g_i^j h_{i'}^j - g_{i'}^j h_i^j |^2 =\sum\limits_{i,i'} \Big[ |g_i^j|^2
|h_{i'}^j|^2 -  (g_i^j h_{i'}^j) (h_i^j g_{i'}^j)^* \Big]\;,
\end{equation}
where $^*$ denotes complex conjugate. After some simple calculations we 
obtain the following relations
\begin{eqnarray}
\sum\limits_{i=1}^{2^{N-1}} |g_i^j|^2 = \langle \psi | \left( \frac
{1+\sigma_z^j}{2} \right) |\psi \rangle, \\ 
\sum\limits_{i=1}^{2^{N-1}} |h_i^j|^2 = \langle \psi | \left( \frac
{1-\sigma_z^j}{2} \right) |\psi \rangle, \\
\sum\limits_{i=1}^{2^{N-1}} g_i^j (h_i^j)^*= \langle \psi | \sigma_-^j 
|\psi\rangle\;,
\end{eqnarray}
and the distance becomes 
$D(l_j(\frac{1}{2})|\psi\rangle , l_j(-\frac{1}{2})|\psi\rangle )=
\frac{1}{4} [1 - \langle
\sigma_z^j \rangle^2 - \langle \sigma_x^j \rangle^2 - \langle \sigma_y^j
\rangle^2]$. Since $Q(\ket{\psi})$ contains a sum over all  qubits (see
Eq. (\ref{meyer})), we finally obtain
\begin{equation} \label{explicit Meyer Wallach}
Q(\ket{\psi}) = 1 - \frac{1}{N} \sum\limits_{j=1}^N \Big[ \langle \sigma_z^j
\rangle^2  + \langle \sigma_x^j \rangle^2 + \langle \sigma_y^j \rangle^2
\Big] = 1 - P_{\fh} (\ket{\psi})\;,
\end{equation}
where $P_{\fh}$ is the purity relative to the local algebra $\fh_{loc}= 
\bigoplus \limits_{j=1}^N \fsu(2)_j$ defined in Section \ref{Nspinexamples}.

\section{Classical limit in the LMG model}
\label{app3}

As we mentioned in Section \ref{lmg}, some critical properties of the
LMG, such as the order parameter or the ground state energy per particle in the
thermodynamic limit, may be obtained using a semi-classical approach. In this
section we sketch a rough analysis of why such approximation is valid (for a 
more extensive analysis, see Ref. \cite{gi2}).

We first define the collective operators 
\begin{equation}
E_{(\sigma,\sigma')} = \sum\limits_{k=1}^N c^\dagger_{k \sigma}c^{\;}_{k \sigma'},
\end{equation}
where $\sigma,\sigma' = \uparrow\mbox{ or }\downarrow$ and the fermionic
operators $c^\dagger_{k \sigma}$ ($c^{\;}_{k \sigma}$) have been defined in
Section \ref{lmg}.  The collective operators satisfy the $\fu(2)$ commutation 
relations (Section \ref{sectionunpurity});  that is
\begin{equation}
\left[ E_{(\sigma,\sigma')}, E_{(\sigma'',\sigma''')} \right ] =
\delta_{\sigma' \sigma''} E_{(\sigma,\sigma''')} -
\delta_{\sigma \sigma'''} E_{(\sigma'',\sigma')} .
\end{equation}

If the number of degenerate levels $N$ is very large, it is useful to  define
the intensive collective operators $\hat{E}_{(\sigma,\sigma')} =
E_{(\sigma,\sigma')}/N$, with commutation relations
\begin{equation}
\left[ \hat{E}_{(\sigma,\sigma')}, \hat{E}_{(\sigma'',\sigma''')} \right ] =
\frac{1}{N} \left ( \delta_{\sigma' \sigma''} \hat{E}_{(\sigma,\sigma''')} -
\delta_{\sigma \sigma'''} \hat{E}_{(\sigma'',\sigma')} \right).
\end{equation}
Therefore, the intensive collective  operators commute in the limit $N
\rightarrow \infty$, they are effectively classical and can be simultaneously 
diagonalized. Similarly, the intensive angular momentum operators 
$J_x/N=(\hat{E}_{(\uparrow,\downarrow)} + \hat{E}_{(\downarrow,\uparrow)})/2$, 
$J_y/N=(\hat{E}_{(\uparrow,\downarrow)} -
\hat{E}_{(\downarrow,\uparrow)})/2i$,  and
$J_z/N=(\hat{E}_{(\uparrow,\uparrow)} - \hat{E}_{(\downarrow,\downarrow)})/2$ 
(with $J_\alpha$ defined in Eqs. (\ref{pseudospin1}), (\ref{pseudospin2}), and
(\ref{pseudospin3}))  commute with each other in the thermodynamic limit, so they
can be thought of as the angular momentum operators of a classical system.

Since the intensive LMG Hamiltonian $H/N$, with $H$ given in Eq.
(\ref{lmghamilt2}), can be written in terms of the intensive angular momentum
operators, it can be regarded as the Hamiltonian describing a classical 
system. The ground state of the LMG model $\ket{g}$ is then an eigenstate of
such  intensive operators when $N \rightarrow \infty$:  $(J_\alpha /N) \ket{g}
= j_\alpha \ket{g}$, $j_\alpha$ being the corresponding eigenvalue. In other
words, when obtaining some expectation values of intensive operators such as
$J_\alpha /N$ or $H/N$ the ground state $\ket{g}$ can be pictured as a
classical angular momentum with fixed coordinates in the 
three-dimensional space (see Fig. \ref{angmomcoord}).
 
This point of view makes it clear why such operators ought to be intensive. 
Otherwise, such a classical limit is not valid and terms of order 1 would 
be important for the calculations of the properties of the LMG model.
Obviously, all these concepts can be extended to more complicated Hamiltonians
such as the extended LMG model, or even Hamiltonians including interactions of
higher orders as in \cite{gi2}.

\end{document}